\documentclass[amsmath,amssymb,amsbsy,prb,twocolumn,superscriptaddress,showpacs]{revtex4-1}

\usepackage{amsmath}
\usepackage{amssymb}
\usepackage[usenames, dvipsnames]{color} 
\usepackage{dcolumn}
\usepackage{graphics}
\usepackage[pdftex]{graphicx}
\usepackage{epstopdf}
\usepackage{leftidx}
\usepackage[breaklinks,colorlinks=true,linkcolor=blue,urlcolor=blue,citecolor=blue]{hyperref}
\usepackage{mathrsfs}
\usepackage{mathtools}
\usepackage{rotating}
\usepackage{empheq}
\usepackage{feynmp}
\begin{document}

\title{Theory of triplon dynamics in the quantum magnet BiCu$_2$PO$_6$}

\author{Kyusung Hwang}
\affiliation{Department of Physics and Centre for Quantum Materials, University of Toronto, Toronto, Ontario M5S 1A7, Canada}

\author{Yong Baek Kim}
\affiliation{Department of Physics and Centre for Quantum Materials, University of Toronto, Toronto, Ontario M5S 1A7, Canada}
\affiliation{Canadian Institute for Advanced Research/Quantum Materials Program, Toronto, Ontario MSG 1Z8, Canada}
\affiliation{School of Physics, Korea Institute for Advanced Study, Seoul 130-722, Korea}

\date{\today}

\begin{abstract}

We provide a theory of triplon dynamics in the valence bond solid ground state of the coupled spin-ladders modelled for BiCu$_2$PO$_6$.
Utilizing the recent high quality neutron scattering data [\href{http://www.nature.com/nphys/journal/v12/n3/full/nphys3566.html}{Nature Physics {\bf 12}, 224 (2016)}] as guides and a theory of interacting triplons via the bond operator formulation, 
we determine a minimal spin Hamiltonian for this system. 
It is shown that the splitting of the low energy triplon modes and the peculiar magnetic field dependence of 
the triplon dispersions can be explained by including substantial Dzyaloshinskii-Moriya and symmetric anisotropic spin interactions.
Taking into account the interactions between triplons and the decay of the triplons to the
two-triplon continuum via anisotropic spin interactions, we provide a theoretical picture that can be used to understand
the main features of the recent neutron scattering experimental data.
\end{abstract}

\maketitle

\section{Introduction}

There has been considerable interest in the emergence of non-trivial paramagnetic ground states of quantum magnets 
with interacting spin $S=1/2$ local moments. Such non-trivial quantum paramagnets would occur in low dimensional 
systems or on geometrically frustrated lattices. Two prominent examples of such quantum paramagnetic ground states are quantum spin
liquids and valence bond solids.\cite{2010_Balents} The determination of the underlying Hamiltonian in such systems, however, has been
a challenge as this often requires the detailed information about the spectra of the elementary excitations as well as
the ground state. 

More specifically, the dispersion of the $S=1/2$ charge-neutral spinon excitations or more accurately the two-spinon
continuum would be an essential information to determine the Hamiltonian for quantum spin
liquids. In the case of the valence bond solids, the triplon dispersions and dynamics would play similar roles.
In two- and three-dimensional quantum magnets, such information is quite scarce and this has remained as one of the 
main challenges both in experimental and theoretical studies of these systems. This is in contrast to magnetically ordered
systems where accurate determination of spin wave spectrum has been around for a long time, which is often used to
infer the minimal spin Hamiltonian.

In this work, we present a theoretical study of triplon dynamics in the valence bond solid ground state of the coupled
spin-ladder system, designed to explain magnetic properties of BiCu$_2$PO$_6$.\cite{2007_Koteswararao,2009_Mentre,2010_Alexander,2010_Tsirlin,2011_Lavarelo,2012_Kohama,2013_Choi,2013_Casola,2013_Sugimoto,2013_Plumb,2015_Plumb} 
In particular, we investigate a possible minimal spin Hamiltonian that is consistent with previous experimental results.
The recent high quality neutron scattering experiments reported in Ref. [\onlinecite{2015_Plumb}] enable us to construct the model using the valuable information on triplon dispersions and two-triplon continuum.
Combining the theoretical results obtained in the bond operator 
formulation\cite{1990_Sachdev,1994_Gopalan,2004_Matsumoto} of the spin model and the spectra of the collective modes measured in the experiments, 
we determine various anisotropic spin interactions. It is found that the low energy properties of the triplons can
only be explained in the presence of substantial Dzyaloshinskii-Moriya and symmetric anisotropic spin interactions.\cite{DM_papers}
It is shown that the interactions between the triplons renormalize the triplon dispersions and more importantly the
anisotropic spin interactions are primarily responsible for the triplon decay into two-triplon continuum.  
Our study of the triplon dynamics in this system provides a useful framework to understand the roles of
various anisotropic spin interactions and presents an opportunity to determine the spin Hamiltonian of
the valence bond solid (VBS) ground states in considerable details.
  
The rest of the paper is organized as follows. We introduce our model Hamiltonian for BiCu$_2$PO$_6$ and 
the VBS ground state in Sec. \ref{sec:model_VBS}. The VBS state and its triplon excitations are described using the bond operator formulation 
in Sec. \ref{sec:quadratic}, where the importance of anisotropic spin interactions is discussed in comparison with 
the experimentally measured triplon dispersions. In Sec. \ref{sec:decay}, it is shown that the anisotropic spin interactions
are responsible for the triplon decay into the two-triplon continuum. Finally our results are summarized 
with an outlook in Sec. \ref{sec:conclusion}.

\section{Model Hamiltonian and VBS state\label{sec:model_VBS}}

We start by describing our model Hamiltonian for BiCu$_2$PO$_6$ and the VBS ground state.

\subsection{Hamiltonian}

\begin{figure}
\centering
\includegraphics[width=\linewidth]{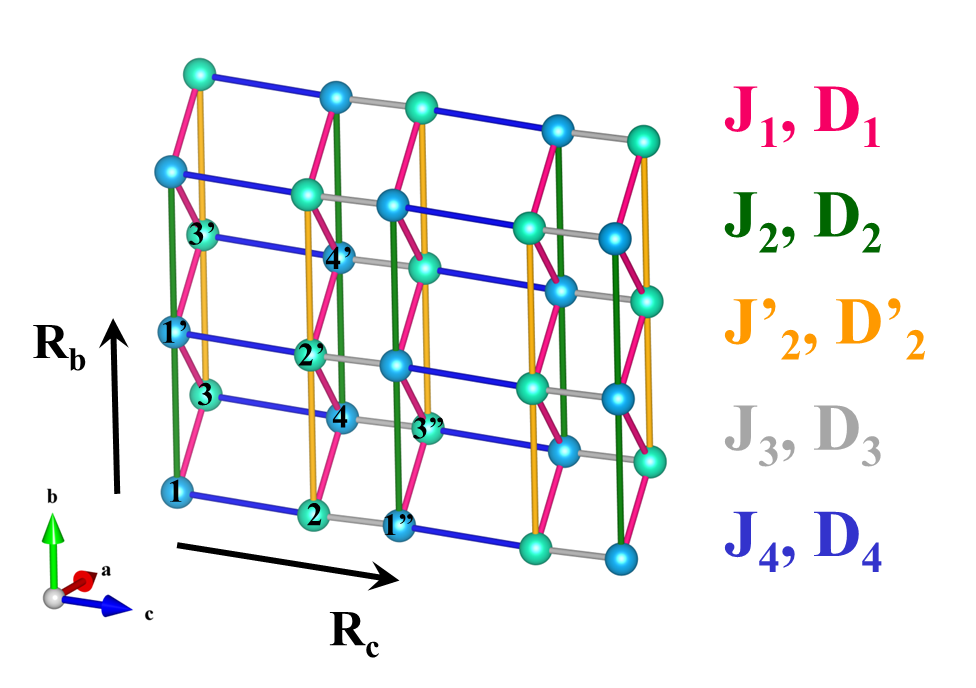}
\caption{(Color online) Lattice structure of the spin-1/2 moments in BiCu$_2$PO$_6$.\label{fig:lattice}}
\end{figure}

We consider the lattice structure in Fig.~\ref{fig:lattice} and introduce the spin model described in Eq.~(\ref{eq:spin_model}) with two types of anisotropic spin interactions, which are the Dzyaloshinskii-Moriya (${\bf D}_{ij}$) and anisotropic \& symmetric ($\Gamma_{ij}^{\mu\nu}$) interactions, as well as the isotropic Heisenberg interactions ($J_{ij}$).
\begin{equation}
 \mathcal{H} = 
 \sum_{i>j} 
 \left(
 J_{ij} {\bf S}_i \cdot {\bf S}_j
 + 
 {\bf D}_{ij} \cdot {\bf S}_i \times {\bf S}_j
 +
 \Gamma_{ij}^{\mu\nu} S_i^{\mu} S_j^{\nu}
 \right),
 \label{eq:spin_model}
\end{equation}
where ${\bf S}_i$ is a $S=1/2$ moment at site $i$,  $\mu,\nu \in \{x,y,z\}$, and summation convention for repeated Greek indices is assumed. We set the $x,y,z$ directions along the crystallographic $a,b,c$ axes (Fig.~\ref{fig:lattice}), respectively. Based on the crystal symmetry $Pnma$ of BiCu$_2$PO$_6$,\cite{2007_Koteswararao,2009_Mentre} there are five independent links as denoted with different colors in the figure. Accordingly, there are five independent Heisenberg interactions: $J_1$ along the zigzag legs (magenta), $J_2$ and $J'_2$ along the straight legs (green and orange, respectively), $J_4$ on the rungs of the ladders (blue), and $J_3$ as the inter-ladder couplings (gray). Distinction between $J_2$ and $J'_2$ arises from the existence of two inequivalent Cu sites (denoted with different colored balls). The antiferromagnetic couplings $J_1$, $J_2$, $J'_2$ form triangular structures on the lattice and generate magnetic frustration in the system. Superexchange pathways are given by Cu-O-Cu for $J_1$ and $J_3$, and Cu-O-O-Cu for $J_2$, $J'_2$, and $J_4$.\cite{2007_Koteswararao,2009_Mentre,2010_Tsirlin} As pointed out in Refs. [\onlinecite{2009_Mentre}] and [\onlinecite{2010_Tsirlin}], the Cu-O-Cu bond angle for the $J_3$ exchange is close to 90$^\circ$ in contrast to the other exchange interactions, implying that $J_3$ can be weak antiferromagnetic or ferromagnetic according to the Goodenough-Kanamori rule.\cite{Goodenough_Kanamori} We assume that the coupling $J_3$ is weak in magnitude compared to the other exchange interactions, $J_1$, $J_2$, $J'_2$, $J_4$, dominated by the antiferromagnetic superexchange. 

The Dzyaloshinskii-Moriya (DM) vectors $\{ {\bf D}_{ij} \}$ can also be determined by the crystal symmetry.\cite{DM_papers} We list the symmetry-constrained DM vectors in Table \ref{tab:indep_coupling} for twelve links in a unit cell. In the table, DM vectors are decomposed along the $a,b,c$ axes: ${\bf D}_{ij}=D_{ij}^a \hat{a}+D_{ij}^b \hat{b}+D_{ij}^c \hat{c}$ with $\hat{a},\hat{b},\hat{c}$ being orthonormal vectors along the crystallographic axes. The minus signs and zero values in the vector components arise from the pseudo-vector nature of ${\bf D}_{ij}$, and {\it mirror} \& {\it inversion} symmetries in the system. 

\begin{table}
\begin{ruledtabular}
\begin{tabular}{ccccc}
 $(i,j)$ & $J_{ij}$ & $D_{ij}^a$ & $D_{ij}^b$ & $D_{ij}^c$
 \\
 \hline
 $(3,1)$ & $J_1$ & $D_1^a$ &  $D_1^b$ &  $D_1^c$
 \\
 $(1',3)$ & $J_1$ & $D_1^a$ &  $-D_1^b$ &  $D_1^c$
 \\
 $(4,2')$ & $J_1$ & $D_1^a$ &  $-D_1^b$ &  $D_1^c$
 \\
 $(2,4)$ & $J_1$ & $D_1^a$ &  $D_1^b$ &  $D_1^c$
 \\
 \\
 $(1,1')$ & $J_2$ & $D_2^a$ &  $0$ &  $D_2^c$
 \\
 $(4',4)$ & $J_2$ & $D_2^a$ &  $0$ &  $D_2^c$
 \\
 \\
 $(2',2)$ & $J'_2$ & ${D'}_2^a$ &  $0$ &  ${D'}_2^c$
 \\
 $(3,3')$ & $J'_2$ & ${D'}_2^a$ &  $0$ &  ${D'}_2^c$
 \\
 \\
 $(1'',2)$ & $J_3$ & $0$ & $D_3^b$ &  $0$
 \\
 $(4,3'')$ & $J_3$ & $0$ & $D_3^b$ &  $0$
 \\
 \\
 $(1,2)$ & $J_4$ & $0$ & $D_4^b$ &  $0$
 \\
 $(4,3)$ & $J_4$ & $0$ & $D_4^b$ &  $0$
 \\
\end{tabular}
\end{ruledtabular}
\caption{Coupling constants $\{J_{ij},{\bf D}_{ij}\}$ determined by the crystal symmetry $Pnma$ of BiCu$_2$PO$_6$. The table lists the coupling constants at twelve interaction links in a unit cell. DM vectors are decomposed along the $a,b,c$ axes, {\it i.e.} ${\bf D}_{ij}=D_{ij}^a \hat{a}+D_{ij}^b \hat{b}+D_{ij}^c \hat{c}$ with $\hat{a},\hat{b},\hat{c}$ being orthonormal vectors along the crystallographic axes. Listed sites, $i,j$, are denoted with numbers in Fig.~\ref{fig:lattice}. The other links on the lattice can be generated by acting lattice translations on the twelve links in the table.
\label{tab:indep_coupling}}
\end{table}

We constrain the coupling constant matrix $\Gamma_{ij}^{\mu\nu}$ of the anisotropic \& symmetric interaction by requiring the following condition:
\begin{equation}
\Gamma_{ij}^{\mu\nu}=\frac{D_{ij}^{\mu} D_{ij}^{\nu}}{2J_{ij}}-\frac{\delta^{\mu\nu}{\bf D}_{ij}^2}{4J_{ij}} \ .
\label{eq:Gamma}
\end{equation}
Here underlying assumption is that the spin exchange interaction is generated by the antiferromagnetic superexchange mechanism. Then, the condition comes from the fact that both of $D$ and $\Gamma$ interactions originate from the spin-orbit coupling in the microscopic Hubbard model (see Ref. [\onlinecite{DM_papers}] or Appendix \ref{app:DM}). As mentioned earlier, $J_3$ is not dominated by the antiferromagnetic superexchange. Hence, Eq.~(\ref{eq:Gamma}) is not applied to the $J_3$ links (gray in Fig.~\ref{fig:lattice}). We find that $D_3$ and $\Gamma_3$ are not essential for describing overall magnetic anisotropies in the system so that we will ignore $D_3$ and $\Gamma_3$ afterwards ($D_3=\Gamma_3=0$).

We will investigate the magnetic field response of the system later. In this case, we consider the Zeeman interaction:
\begin{equation}
 \mathcal{H}_Z = - g \mu_B {\bf H} \cdot \sum_i {\bf S}_i,
 \label{eq:Zeeman}
\end{equation}
with ${\bf H}$ being the magnetic field. In principle, we could consider symmetry-allowed $g$-tensors for two inequivalent Cu$^{2+}$ ions to allow anisotropy in the Zeeman interaction as well. However, this will introduce more parameters needed to be determined, 
rendering the theory more complicated. For simplicity, we assume an isotropic $g$-factor with $g=2$.

\subsection{Valence bond solid}

\begin{figure}
\centering
\includegraphics[width=0.75\linewidth]{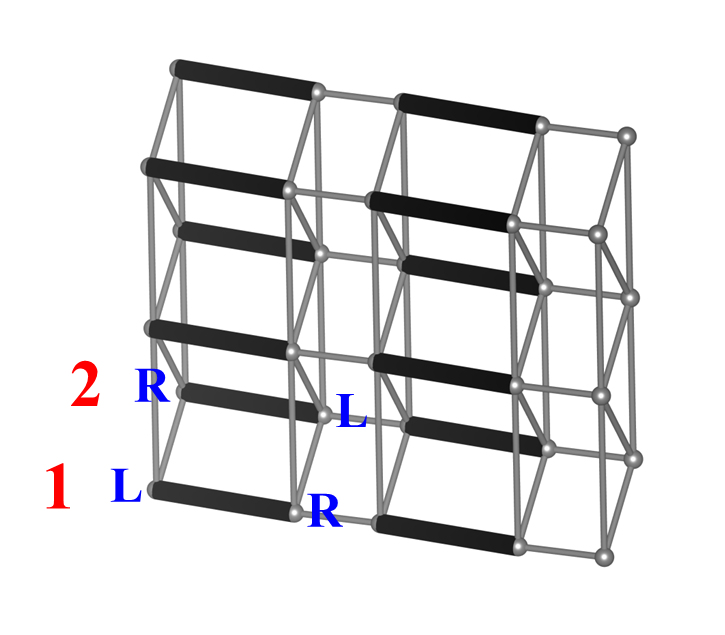}
\caption{(Color online) Dimer covering of the rung-VBS phase. Dimers (thick black lines) denote the valence bonds formed along the link. The figure also shows the convention for the dimer index (1,2) and the spin index (L,R) within a dimer in the bond operator theory.
\label{fig:VBS}}
\end{figure}

Now we discuss the valence bond solid phase as the ground state of BiCu$_2$PO$_6$. The VBS phase is depicted in Fig. \ref{fig:VBS}. Here, the valence bonds are formed at the $J_4$-links or the rungs of the spin ladders (denoted with thick black lines). We call this phase the rung-VBS in this paper. 

The existence of  a VBS phase in BiCu$_2$PO$_6$ has been hinted through earlier studies. A finite spin gap and elementary spin-1 excitations in the compound are evidences for a VBS state. The finite spin gap was observed in various experiments such as the heat capacity, magnetic susceptibility, NMR, and neutron scattering.\cite{2007_Koteswararao,2009_Mentre,2010_Alexander,2010_Tsirlin,2013_Plumb} The spin-1 excitations have been detected in recent inelastic neutron scattering experiments.\cite{2013_Plumb,2015_Plumb} The spin-1 character of the excitations was inferred by investigating their behaviours under external magnetic fields. The structure of the VBS state was investigated in the elaborate work by Tsirlin {\it et al.}\cite{2010_Tsirlin} They constructed a Heisenberg spin ladder model and studied it by using various numerical techniques and experimental informations. It was shown that the ground state of the Heisenberg spin ladder model has strong bond strength $\langle {\bf S}_i\cdot{\bf S}_j\rangle$ at the rungs in the exact diagonalization studies. This numerical result suggests that the rung-VBS phase arises in their model, where they didn't consider anisotropic spin interactions. The spin model in Eq. (\ref{eq:spin_model}) is a generalization of the Heisenberg spin ladder model with anisotropic spin interactions, which turns out to be extremely important to describe the neutron scattering data.

In the next section, we show that the model in Eq. (\ref{eq:spin_model}) provides an excellent description of the triplon excitations seen in the scattering experiments on BiCu$_2$PO$_6$. In the following, we will describe the triplon excitations in the bond operator theory developed for the rung-VBS phase.

\section{Bond operator formulation\label{sec:quadratic}}

Bond operator theory\cite{1990_Sachdev,1994_Gopalan,2004_Matsumoto} is a useful framework for describing a valence bond solid phase and its spin-1 triplon excitations. The theory is built upon the bond operator representation of the spin operators ${\bf S}_{L,R}$ forming a valence bond.
\begin{equation}
\label{eq:bond_op}
\begin{array}{c}
 S_L^{\alpha} = \frac{1}{2} \left( s^{\dagger}t_{\alpha} + t_{\alpha}^{\dagger}s - i\epsilon_{\alpha\beta\gamma}t_{\beta}^{\dagger}t_{\gamma} \right),
 \\
 S_R^{\alpha} = \frac{1}{2} \left( - s^{\dagger}t_{\alpha} - t_{\alpha}^{\dagger}s - i\epsilon_{\alpha\beta\gamma}t_{\beta}^{\dagger}t_{\gamma} \right),
\end{array}
\end{equation}
where $\alpha,\beta,\gamma\in\{x,y,z\}$, and $\epsilon_{\alpha\beta\gamma}$ is the totally antisymmetric tensor.
The bond operators $s^{\dagger}$ and $t_{\alpha}^{\dagger}$ create the spin-singlet and spin-triplet states of ${\bf S}_{L,R}$, respectively, and follow the boson statistics. In order to keep the physical Hilbert space consisting of the four states (the singlet and triplet), the hardcore constraint should be imposed: $s^{\dagger}s + t_{\alpha}^{\dagger} t_{\alpha} = 1$. 

We express the original spin Hamiltonian Eq. (\ref{eq:spin_model}) in terms of the bond operators by using the representation Eq. (\ref{eq:bond_op}) with the dimer covering for the rung-VBS state in Fig. \ref{fig:VBS}. In the resulting bond operator Hamiltonian, the rung-VBS state can be described by condensing the $s$-bosons at all the dimers: $\langle s \rangle =\langle s^{\dagger} \rangle =\bar{s}$. The hardcore constraint is incorporated at each dimer with the Lagrange multiplier $\mu$ as $-\mu(s^{\dagger}s + t_{\alpha}^{\dagger} t_{\alpha} - 1)$. Exploiting the crystal symmetry of BiCu$_2$PO$_6$, we set the parameters $\{\bar{s},\mu\}$ to be uniform across all the dimers. Then, we end up with the following form of the bond operator Hamiltonian.
\begin{equation}
 \mathcal{H} + \mathcal{H}_Z = \mathcal{H}_{quad} + \mathcal{H}_{cubic} + \mathcal{H}_{quartic}.
 \label{eq:H_BOT}
\end{equation}
Here, the Hamiltonian is arranged according to the order of the $t$-boson operator. $\mathcal{H}_{quad}$ consists of the quadratic terms like $t^{\dagger} t$, $t^{\dagger} t^{\dagger}$, and their Hermitian conjugates. $\mathcal{H}_{cubic}$ contains the cubic terms $t^{\dagger} t^{\dagger} t$ and $t^{\dagger} t t$. $\mathcal{H}_{quartic}$ has the quartic terms of the form $t^{\dagger} t t^{\dagger} t$. In the above expression, we also included the Zeeman interaction [Eq. (\ref{eq:Zeeman})], which only has quadratic terms since possible linear terms are cancelled.

Below, we develop a simple quadratic bond operator theory by keeping only the quadratic part $\mathcal{H}_{quad}$ in the Hamiltonian. Via this quadratic theory, we describe the low energy triplon excitations around the spin gap observed in experiments. We will consider higher order interactions later in this paper.

\subsection{Quadratic Hamiltonian\label{sec:quadratic_theory}}

The quadratic bond operator Hamiltonian has the following form in the momentum space after the Fourier transformation.
\begin{equation}
  \mathcal{H}_{quad} = 
 E_o 
 +
 \frac{1}{2}\sum_{\bf k} \Lambda^{\dagger}({\bf k}) \boldsymbol{\mathcal{M}}({\bf k}) \Lambda({\bf k}),
\end{equation}
where $E_o$ is a function of $\bar{s}$ and $\mu$, and 
\begin{equation}
\Lambda({\bf k})=
\left(
 \begin{array}{c}
 {\bf t}_{1}({\bf k})
 \\
 {\bf t}_{2}({\bf k})
 \\
 {\bf t}_{1}^{\dagger}(-{\bf k})
 \\
 {\bf t}_{2}^{\dagger}(-{\bf k})
 \end{array}
 \right)
\end{equation}
with
\begin{equation}
{\bf t}_{1}({\bf k})
=
\left(
 \begin{array}{c}
 t_{1x}({\bf k})
 \\
 t_{1y}({\bf k})
 \\
 t_{1z}({\bf k})
 \end{array}
 \right),
 ~~~
{\bf t}_{1}^{\dagger}(-{\bf k})
=
\left(
 \begin{array}{c}
 t_{1x}^{\dagger}(-{\bf k})
 \\
 t_{1y}^{\dagger}(-{\bf k})
 \\
 t_{1z}^{\dagger}(-{\bf k})
 \end{array}
 \right),
\end{equation}
and similarly for ${\bf t}_{2}({\bf k})$ and ${\bf t}_{2}^{\dagger}(-{\bf k})$. Here, the subscripts, 1 and 2, in the $t$-operators indicate the two dimers in a unit cell (shown in Fig. \ref{fig:VBS}). As mentioned earlier, the $x,y,z$ directions are taken parallel to the crystallographic $a,b,c$ axes. The $J$, $D$, $\Gamma$, and $H$ interactions in the original spin Hamiltonian are transformed to the triplon hopping and pairing amplitudes contained in the 12$\times$12 matrix $\boldsymbol{\mathcal{M}}({\bf k})$. The quadratic Hamiltonian has two notable features: (i) its dependences on $J_2$ and $J'_2$ appear only through the sum $J_2+J_2'$, and (ii) the $D_1$ interactions cancel each other at the quadratic level without any contribution to $\mathcal{H}_{quad}$. These features will be discussed again later. The detailed form of $\mathcal{H}_{quad}$ is provided in Appendix \ref{app:matrix}.

The quadratic Hamiltonian is diagonalized via the Bogoliubov transformation leading to the following form:
\begin{equation}
 \mathcal{H}_{quad} = E_{gs} + \sum_{\bf k} \sum_{n=1}^{6} \omega_n({\bf k}) \gamma_{n}^{\dagger}({\bf k}) \gamma_{n}({\bf k}).
\end{equation}
Here, $\gamma_{n}^{\dagger}({\bf k})$ is the Bogoliubov quasiparticle operator or the {\it triplon} with the excitation energy $\omega_n({\bf k})$.
The constant term corresponds to the ground state energy, $E_{gs}=\langle \mathcal{H}_{quad} \rangle$.
With this setting, the parameters $\bar{s}$ and $\mu$ for the ground state are determined by the saddle point equations:
\begin{equation}
 \frac{\partial \langle \mathcal{H}_{quad} \rangle}{\partial \bar{s}} = 0 ,
 ~~~~~
 \frac{\partial \langle \mathcal{H}_{quad} \rangle}{\partial \mu} = 0 .
 \label{eq:saddle-point-eq}
\end{equation}
In the ground state, we also compute the magnetization, ${\bf M}$, under nonzero magnetic fields as follows.
\begin{equation}
{\bf M} = \frac{1}{N} \langle g \mu_B \sum_i {\bf S}_i \rangle ,
\label{eq:magnetization}
\end{equation}
with $N$ being the number of the spin moments.

Before discussing the results of the quadratic theory, we comment on our conventions about the Brillouin zone and momentum vectors. For direct comparisons of our theoretical computations with experimental results, we use the extended zone scheme for the Brillouin zone and denote momentum vectors in reciprocal lattice unit as ${\bf q}=(h,k,l)$, which means ${\bf q}=h{\bf G_a}+k{\bf G_b}+l{\bf G_c}$ with ${\bf G_{a,b,c}}$ being the reciprocal lattice vectors corresponding to the lattice vectors ${\bf R_{a,b,c}}$ along the $a,b,c$ axes, repsectively.

\subsection{Triplon dispersions\label{sec:triplon_dispersions}}

\begin{figure}[t]
\centering
\includegraphics[width=0.7\linewidth,angle=270]{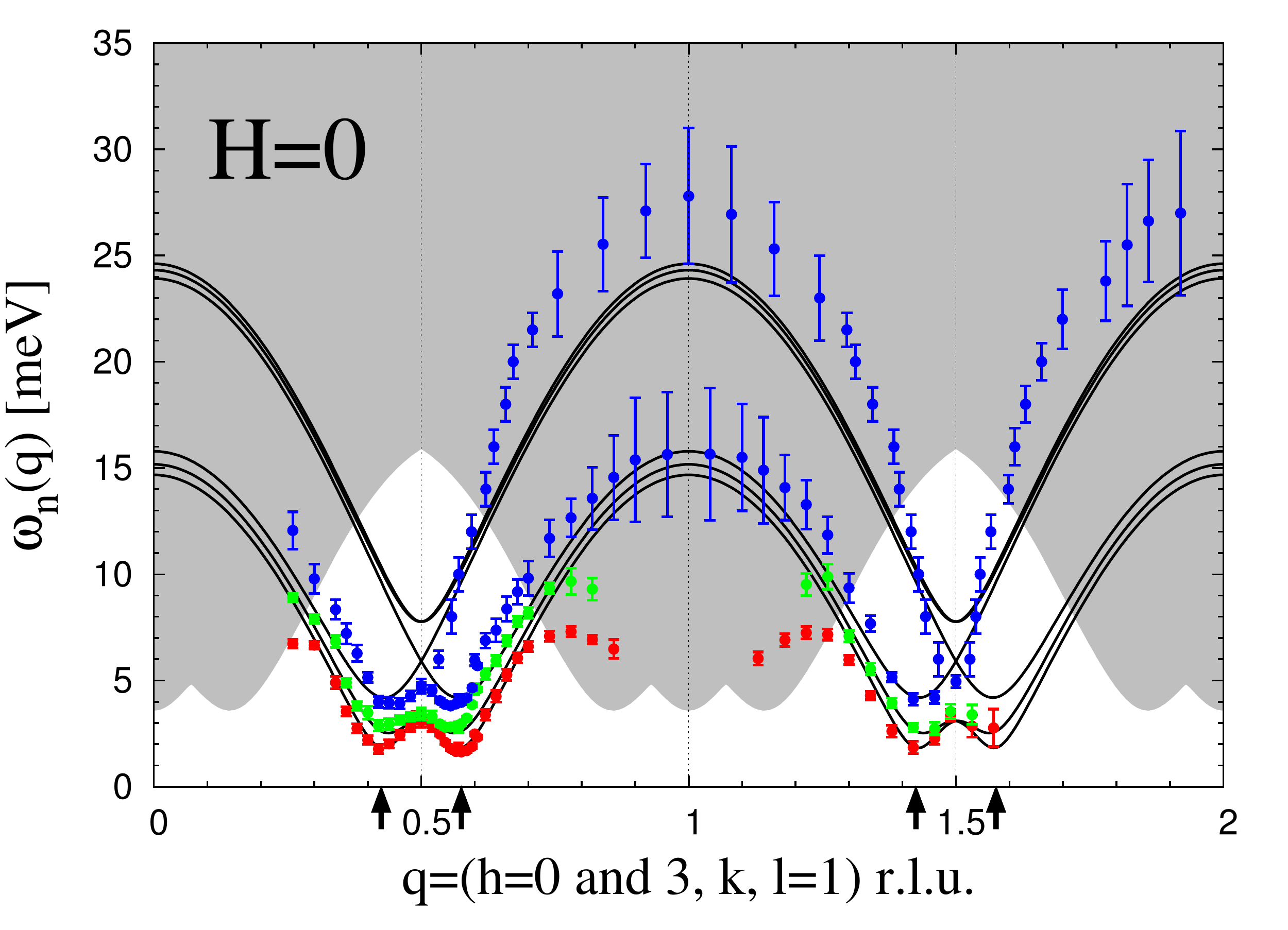}
\caption{(Color online) Triplon energy dispersions, $\omega_n({\bf q})$, in the quadratic bond operator theory. The dispersions are compared with the neutron scattering results in Ref. [\onlinecite{2015_Plumb}] (denoted wth color dots). The spin gap positions, $\pm{\bf q}^{\star}=\pm (0,0.425,0)$, are marked by the arrows. The gray region indicates the multi-triplon continuum computed with Eq. (\ref{eq:two-triplon}).
Error bars in the neutron scattering data represent the energy widths (or decay rates) of the measured quasiparticle peaks.
\label{fig:quad_band}}
\end{figure}

The triplon dispersions obtained in the quadratic theory can be fitted with the experimental results\cite{2015_Plumb} by controlling the coupling constants $\{J,D,\Gamma\}$. Figure \ref{fig:quad_band} shows the best fit (black lines) with the neutron scattering data (color dots) along the momentum direction ${\bf q}=(h=0~\textup{and}~3,~k,~l=1)$, yielding the following set of the coupling constants.
\begin{equation}
\begin{array}{l}
J_1=J_2=J'_2=J_4=8~\textup{meV},
\\
J_3=0.2J_1,
\\
\Gamma_1^{aa}=-\Gamma_1^{bb}=0.039J_1,
\\
\Gamma_1^{ab}=\Gamma_1^{ba}=0.135J_1,
\\
(D_1^a=0.6J_1, ~ D_1^b=0.45J_1).
\end{array}
\label{eq:coupling_const}
\end{equation}
The other coupling constants not shown here are set to zero since they are found to be irrelevant for describing essential features of the neutron scattering data.
As mentioned earlier, the $D_1$ couplings cancel each other in the quadratic Hamiltonian $\mathcal{H}_{quad}$ whereas the $\Gamma_1$ couplings survive at the quadratic level. Hence, we control the $\Gamma_1$ couplings instead of the $D_1$ couplings. The $D_1$ couplings in Eq. (\ref{eq:coupling_const}) are the values obtained from the $\Gamma_1$ couplings and  the relationship Eq. (\ref{eq:Gamma}). As will be shown later, the $D_1$ interactions appear in the cubic terms of $\mathcal{H}_{cubic}$. Their effects on the triplon excitations will be investigated in a later part of this paper. Further discussions on the parameter regime of Eq. (\ref{eq:coupling_const}) are provided in Appendix \ref{app:fitting}.

The quadratic theory with the coupling constants in Eq. (\ref{eq:coupling_const}) yields {\it six nondegenerate}, triplon dispersions (see Fig. \ref{fig:quad_band}). It is due to the fact that the anisotropic and symmetric couplings, $\Gamma_1$, completely break the SO(3) spin rotation symmetry existing at the level of Heisenberg model, and there are two dimers in a unit cell. The spin gap (minimum excitation energy) occurs at the incommensurate momentum positions: $\pm{\bf q}^{\star}=\pm (0,0.425,0)$ and their equivalent momenta translated by reciprocal lattice vectors (denoted with arrows in the figure). We find that the lowest three dispersions are in good agreement with the neutron scattering results around the spin gap. Among the other higher-energy three dispersions, only one of them is experimentally observed and qualitatively consistent with the theoretical result (see the highest line of blue dots). 

In regions away from the spin gap, however, the quadratic theory cannot fully explain the results from the experimental measurements. To be specific, inside the gray region, the lowest dispersion bends downward (red), and the lowest two triplon modes decay at certain momenta (red and green). These features are believed to be the effects of triplon interactions coming from, for example, the cubic terms generated by the $D_1$ interactions. These effects will be discussed later. At the moment, we focus on the low energy part around the spin gap (below the gray region) and see if the quadratic theory provides a satisfactory description of the low energy spin dynamics in BiCu$_2$PO$_6$. As we have already seen, the theory is in good agreement with the neutron scattering results in the low energy region. This fact supports the idea that the spin excitations observed in the neutron scattering are triplon excitations, confirming the rung-VBS state in BiCu$_2$PO$_6$. Below we provide more evidences for this conclusion.

\subsection{Magnetic field response\label{sec:triplon_magnetic_behaviors}}

\begin{figure*}[t!]
\centering
\includegraphics[width=0.3\linewidth,angle=270]{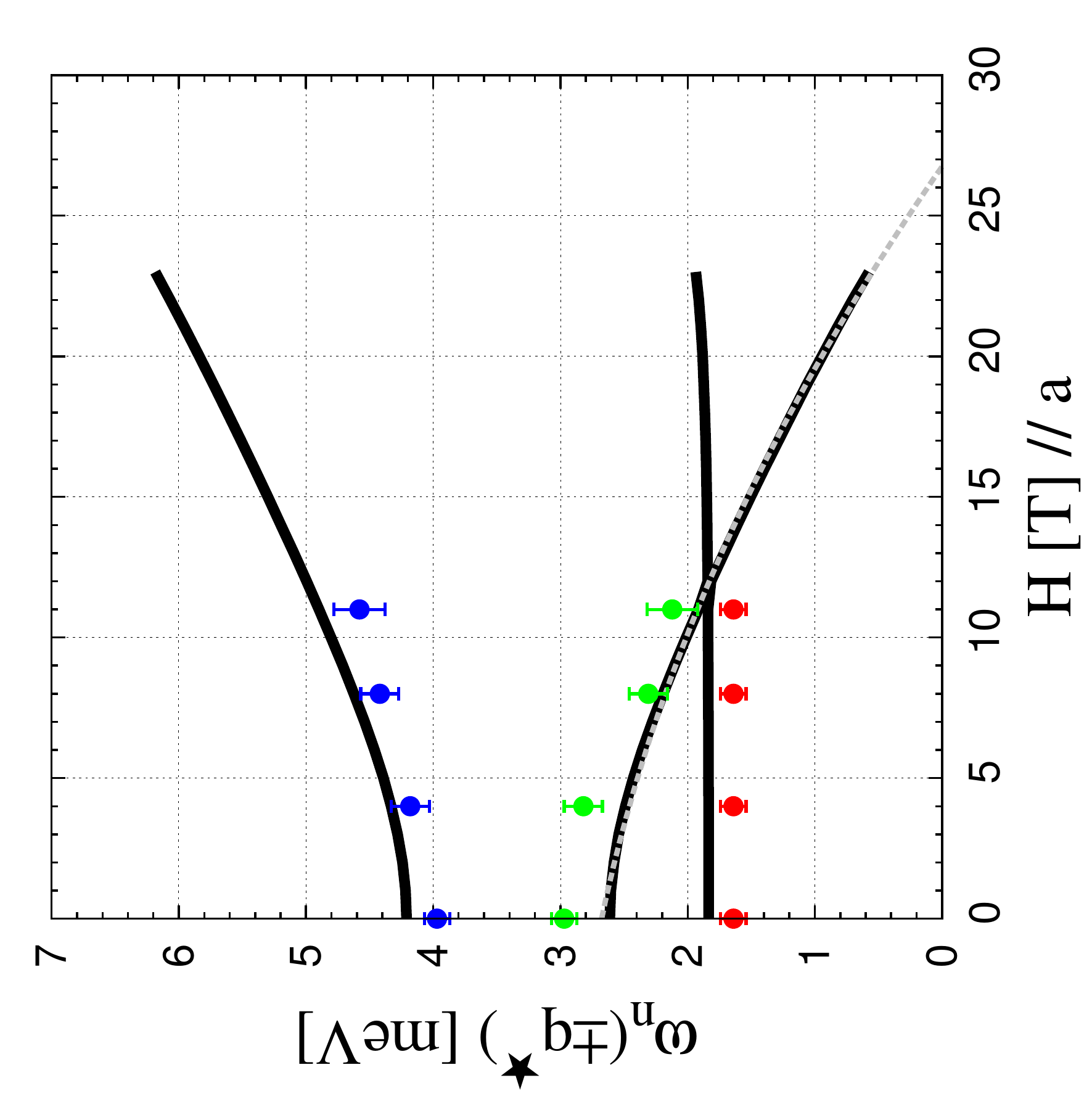}
\includegraphics[width=0.3\linewidth,angle=270]{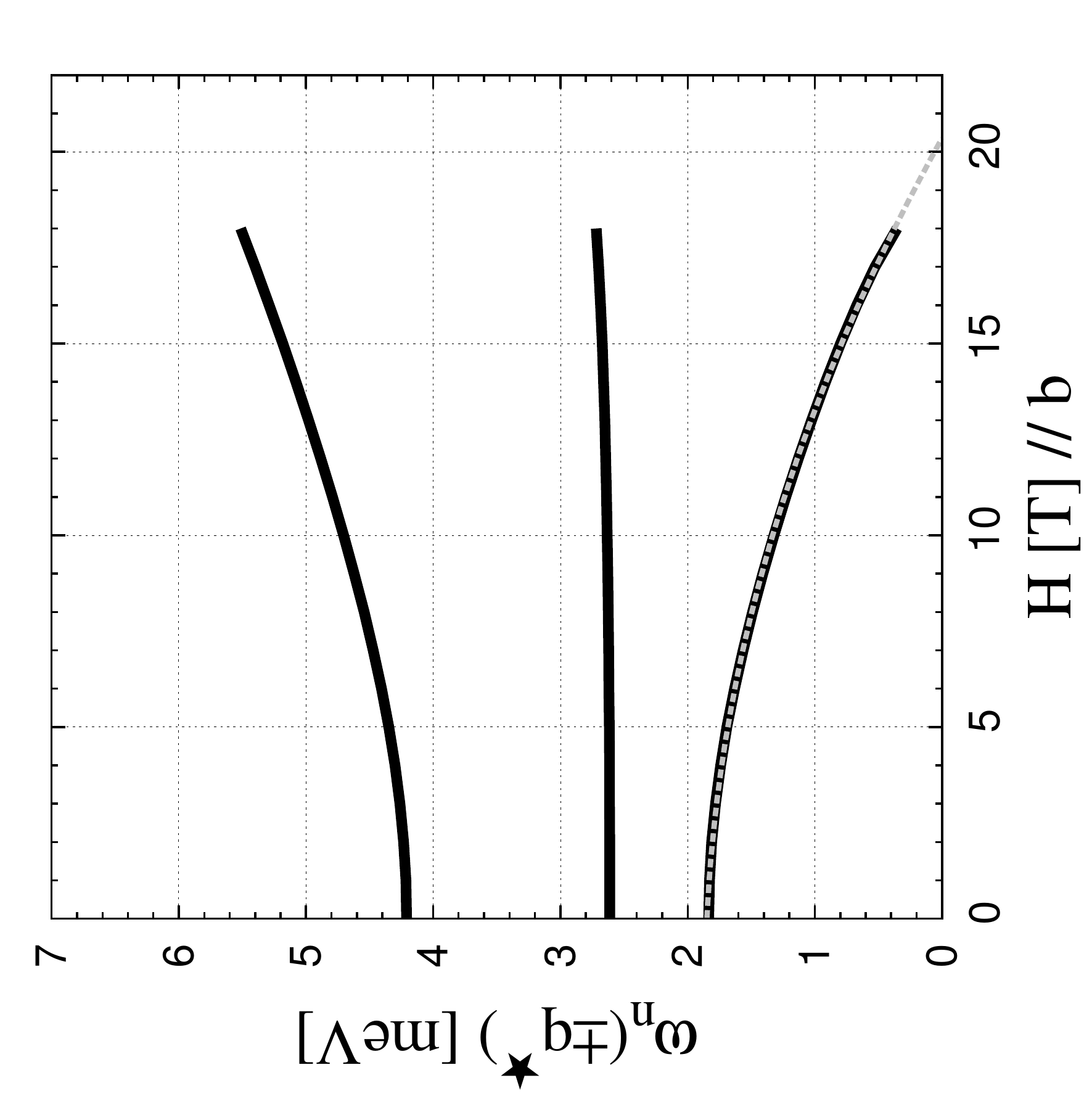}
\includegraphics[width=0.3\linewidth,angle=270]{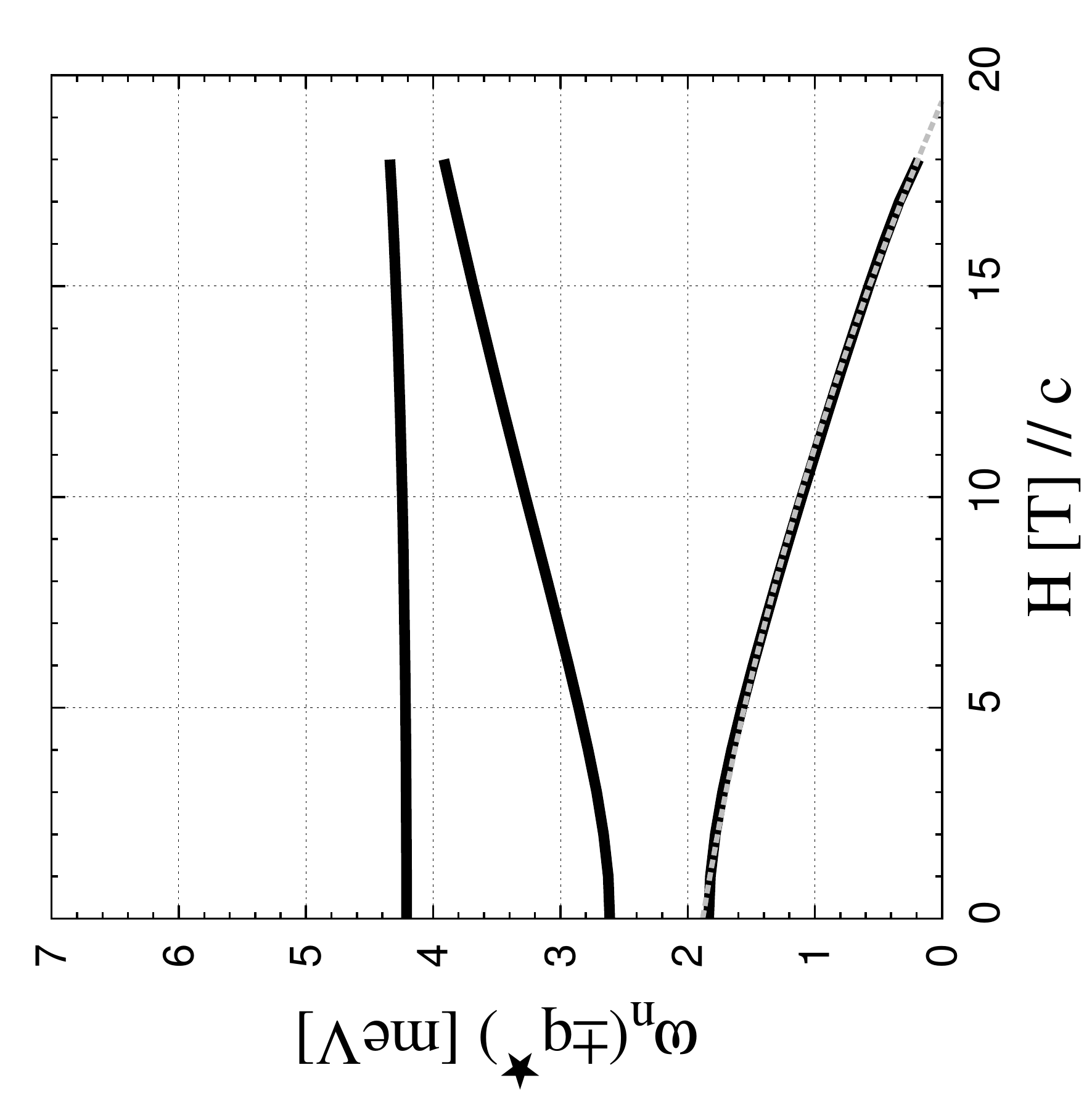}
\caption{ (Color online) Magnetic field dependence of the triplon excitations in the quadratic bond operator theory. The three plots show the triplon excitation energies at the spin gap positions, $\omega_n(\pm{\bf q}^{\star})$, as a function of the magnetic field, $H$, for the field directions along the $a,b,c$ axes, respectively. In the case of the field along the $a$ axis (the left), the theoretical results are compared with the neutron scattering data in Ref. [\onlinecite{2015_Plumb}] (colored dots).  For an estimation of the critical field, $H_c$, in each plot the theoretical results are extrapolated with a functional form ($c_0 + c_1 H + c_2 H^2$) as depicted with a dashed line.\cite{extrapolation}
\label{fig:Zeeman}
}
\end{figure*}

As another mean to check that the observed excitation modes are the triplons in the rung-VBS phase, their field responses can be examined in the theory and compared with the experimental data. Turning on the Zeeman interaction in the quadratic bond operator Hamiltonian, we compute the triplon excitation spectrum as a function of the magnetic field. The obtained spectra are plotted in Fig. \ref{fig:Zeeman} at the spin gap positions, $\pm{\bf q}^{\star}$. Existing experimental data\cite{2015_Plumb} is only for the fields along the $a$ axis, which is also denoted in the figure with color dots. One can find that the theoretical result is consistent with the experimental data both qualitatively and quantitatively (left panel). For example, it was observed in experiments that the lowest mode (red) is almost not reacting to the magnetic field whereas the other two higher energy modes are moving downward (green) and upward (blue), respectively. These behaviours are well captured in the theoretical result. The observed magnetic field response also tells us that the triplon modes do not possess any degeneracy as predicted in the theory.

Notice that, in general, the magnetic field dependence of the triplon dispersion is not linear in magnetic field, especially for the two higher energy triplon modes
(green and blue dots in Fig. \ref{fig:Zeeman}). Indeed the magnetic field dependence of the two higher energy modes follows 
$c_0 + c_1 H + c_2 H^2$ behaviour. This is due to the fact that the triplon modes do not possess well-defined spin quantum number 
(${\bf S} \cdot \hat{H}$, spin component along the field direction) as a result of magnetic anisotropies. Instead the usual spin quantum numbers
(${\bf S} \cdot \hat{H}=+1,0,-1$) in the spin isotropic case are mixed in the triplon modes. The non-linear magnetic field dependence 
can also be found in other field directions.

It can be seen that certain triplon modes have almost constant energy in magnetic field. For instance, when the magnetic field is applied along 
the $a$ axis, the energy of the lowest energy triplon mode is basically constant (red dots in Fig.  \ref{fig:Zeeman}), implying that the spin 
character of this mode is dominated by the quantum number ${\bf S} \cdot \hat{a}=0$. Similar behavior is also found in the 
second/third lowest mode under the magnetic field along $b$/$c$ axis (see the middle and right panels in Fig. \ref{fig:Zeeman}). 
Accordingly, we can characterize three triplon modes approximately as the ${\bf S} \cdot \hat{a}=0$, ${\bf S} \cdot \hat{b}=0$, ${\bf S} \cdot \hat{c}=0$ 
states from the lowest to the highest energy modes (the spin characters can be directly identified by taking the spin projections of the triplon mode eigenvectors). 
In other words, three triplon modes have their own approximate spin quantization axes. If the applied magnetic field is not along the quantization axis, triplon modes 
follow non-linear magnetic field dependence and can be characterized by mixed spin states.

\subsection{Magnetization and critical field\label{sec:magnetization}}

\begin{figure}[b]
\centering
\includegraphics[width=0.7\linewidth,angle=270]{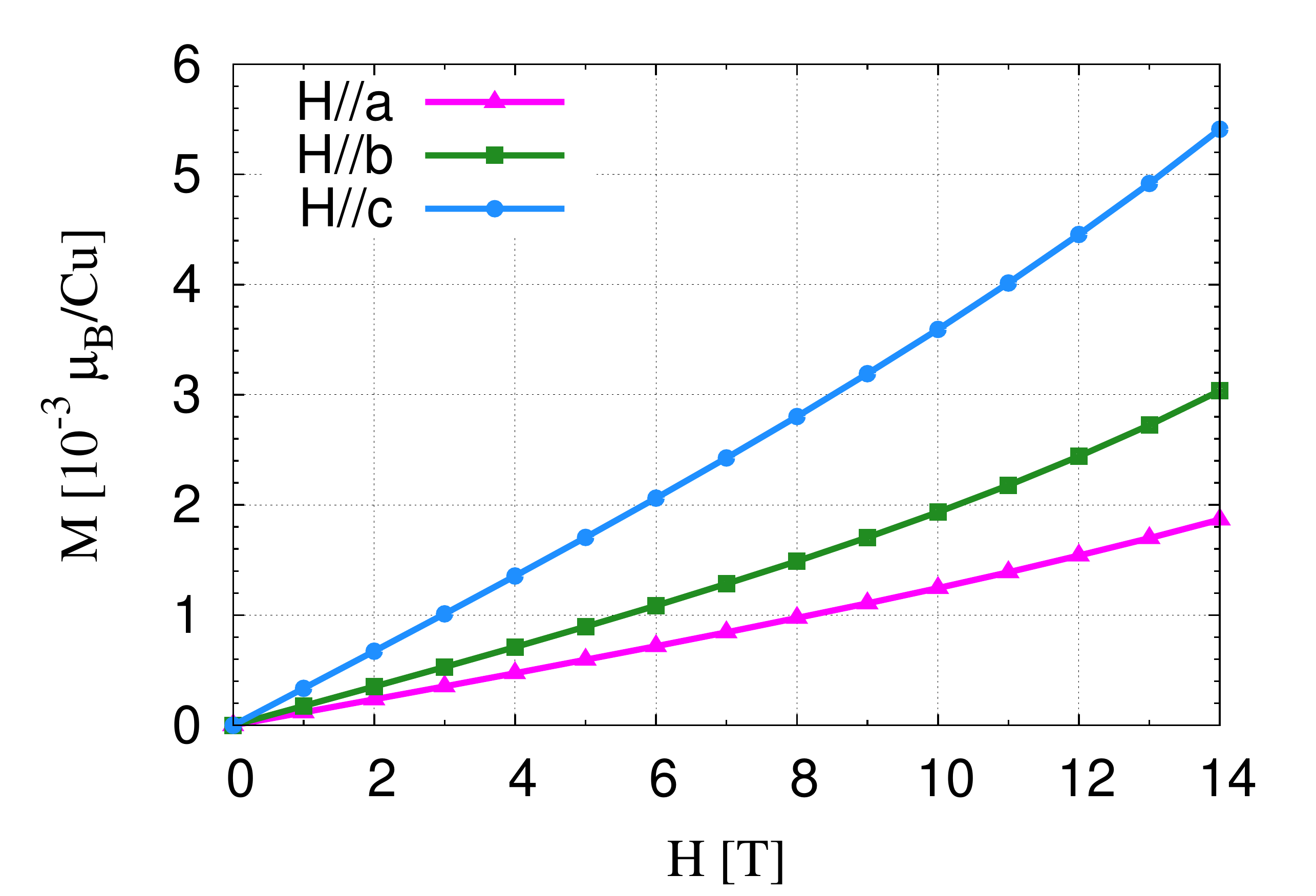}
\caption{(Color online) Magnetization, $M$, as a function of magnetic field, $H$, obtained in the theory. 
The magnetization is computed with Eq. (\ref{eq:magnetization}) for the magnetic fields along the $a,b,c$ axes. The figure shows the
the magnetization components along the field directions. Other components perpendicular to the field are zero. 
\label{fig:Magnetization}}
\end{figure}

We now discuss the magnetization process of the system. Recent high magnetic field experiments in Ref. [\onlinecite{2012_Kohama}] show that BiCu$_2$PO$_6$ undergoes a cascade of field-induced phase transitions with anisotropic magnetic responses to different field directions. Upon increasing the field, the magnetization monotonically increases with different slopes depending on the field direction until the system reaches the transition where the spin gap is closed. The average slope or susceptibility, ${\chi}_{avg}=\frac{\Delta M}{\Delta H}$, has the sequence of ${\chi}_{avg}^c > {\chi}_{avg}^b > {\chi}_{avg}^a$ and concomitantly the critical field, $H_c$, has the opposite sequence of $H_c^a ~(= 23 \textup{T}) > H_c^b ~(= 21 \textup{T}) > H_c^c ~(= 20 \textup{T})$, where the superscripts denote the applied field direction. 

The above experimental features can be explained in the quadratic bond operator theory. Figure \ref{fig:Magnetization} shows the magnetizations predicted from the theory, which are consistent with the susceptibility sequence pattern observed in the experiments. On the other hand, the critical fields can be read from the previous triplon energy plots in Fig. \ref{fig:Zeeman}. Extrapolating the triplon spectra (gray dashed lines), we find that $H_c^a ~(= 27 \textup{T}) > H_c^b ~(= 20 \textup{T}) > H_c^c ~(= 19 \textup{T})$ in the quadratic theory. Although the value of $H_c^a$ is a bit larger than the measured value, the right trend in the sequence of the critical fields is well captured in the theory.

\subsection{Importance of triplon interactions\label{sec:triplon_int_imp}}

In the previous sections, we have observed remarkable consistency between the quadratic bond operator theory and experiment in the low energy regions. This fact implies that BiCu$_2$PO$_6$ has the rung-VBS ground state with the triplons as the elementary excitations. Moreover, the theory tells us that the couplings in Eq. (\ref{eq:coupling_const}) are minimal interactions. 

On the other hand, the experimental data deviates from the theoretical calculations in the gray region of Fig. \ref{fig:quad_band} with the following features: (i) downward bending of the lowest triplon dispersion (red), (ii) abrupt decays of the lowest two triplon modes at certain momenta (red and green), and (iii) broadened energy width or increased decay rate in the third lowest dispersion (blue). Notice that the gray region in the figure denotes multi-triplon continuum with the lower boundary computed as the minimum energy of two-triplon excitations:
\begin{equation}
 E_2({\bf q}) = \underset{{\bf k},m,n}{\textup{min}} [ \omega_m ({\bf q}-{\bf k}) + \omega_n ({\bf k}) ],
 \label{eq:two-triplon}
\end{equation}
where $\omega_n ({\bf k})$ is the single-triplon dispersion in the quadratic theory with $m,n$ being band indices. This fact proposes a picture that the triplons are strongly interacting within the triplon continuum region so that they lose their prominence as quasiparticle modes inside the region. Thus, we need to consider the interactions between triplons to capture triplon decay processes.

\section{Effects of triplon interactions\label{sec:decay}}

Now we consider the influence of triplon interactions on the triplon dynamics in BiCu$_2$PO$_6$. Two main effects are expected from the triplon interactions: triplon energy renormalization and decay.\cite{2013_Zhitomirsky,2006_Zhitomirsky} These two effects are closely related with the triplon phenomenology observed in the neutron scattering experiments (substantial energy splittings around the spin gap and decay phenomena inside the triplon continuum).

The triplon interactions are taken into account by extending the previous quadratic bond operator theory. For the spin Hamiltonian [Eq. (\ref{eq:spin_model})] with the couplings in Eq. (\ref{eq:coupling_const}), we arrange the corresponding bond operator Hamiltonian in the following way:
\begin{equation}
 \mathcal{H} = \mathcal{H}_{quad}[\Gamma_1] + \mathcal{H}_{cubic}[D_1] + \mathcal{H}_{quartic}[D_1,\Gamma_1],
\end{equation}
where we have denoted the dependence on the anisotropic couplings in the square brackets. We will take the mean-field approximations for $\mathcal{H}_{quartic}$ and use the cubic interactions, $\mathcal{H}_{cubic}$, to describe the triplon decay processes. In this approach, the triplon modes and their decays are identified via the peak structures in the spectral weight function of the triplon Green's function. This approach reveals that the $D_1$ couplings have noticeable contributions on the energy splittings around the spin gap as well as the decay phenomena inside the triplon continuum. Readers interested in the results rather than the calculational details are advised to directly move to Sec. \ref{sec:result}.

\subsection{Mean-field approximations for $\mathcal{H}_{quartic}$}

The quartic terms in $\mathcal{H}_{quartic}$ provide two-body scatterings of the $t$-bosons. We include the two-body interactions in the mean-field approximations, leading to the following mean-field Hamiltonian:
\begin{equation}
 \mathcal{H}_{quad}+\mathcal{H}_{quartic} \rightarrow \mathcal{H}_{MF}.
\end{equation}
For the mean-field decouplings, we employ particle-particle ($Q=\langle t t \rangle$) and particle-hole ($P=\langle t t^{\dagger} \rangle$) channels. Details of the decoupling scheme are explained in Appendix \ref{app:MF}.

\subsection{Triplon decay channels of $\mathcal{H}_{cubic}$\label{sec:decay_channel}}

The cubic terms in $\mathcal{H}_{cubic}$ provide decay and fusion processes of the $t$-bosons ($t^{\dagger} \rightleftharpoons t^{\dagger} t^{\dagger}$). The processes are induced by the cubic terms from the $D_1$ and $J_3$ couplings whereas the other couplings, $J_1$, $J_2(=J'_2)$, $J_4$, and $\Gamma_1$, do not have such cubic terms because of symmetry reasons. More details about the existence of the cubic terms and the associated symmetry are provided in Appendix \ref{app:decay}.

We will investigate the effects of the cubic interactions via the Green's function approach.\cite{2013_Zhitomirsky,2006_Zhitomirsky,2003_Fetter} In this approach, the triplons determined from the mean-field Hamitonian, $\mathcal{H}_{MF}$, are taken as the bare particles. We express the interaction part $\mathcal{H}_{cubic}$ in terms of the bare triplons ($\gamma$) from $\mathcal{H}_{MF}$, which leads to the following form:
\begin{widetext}
\begin{eqnarray}
 \mathcal{H}_{cubic}
 &=&
 \sum_{l,m,n=1}^{6} \sum_{{\bf k}+{\bf p}-{\bf q}={\bf 0}} \frac{1}{2!} Y_2(k_l,p_m;q_n) \gamma^{\dagger}(k_l) \gamma^{\dagger}(p_m) \gamma(q_n) + \textup{H.c.}
 \nonumber\\
 &+&
 \sum_{l,m,n=1}^{6} \sum_{{\bf k}+{\bf p}+{\bf q}={\bf 0}} \frac{1}{3!} Y_3(k_l,p_m,q_n) \gamma^{\dagger}(k_l) \gamma^{\dagger}(p_m) \gamma^{\dagger}(q_n) + \textup{H.c.}
 \label{eq:cubic}
\end{eqnarray}
\end{widetext}
Here, we use the shorthand notations: $k_l=({\bf k},l)$, and so forth for $p_m$, $q_n$. In addition to the decay and fusion terms for the triplons ($Y_2$ and $Y_2^*$), we have the source and sink terms for the $\gamma$-triplons ($Y_3$ and $Y_3^*$) in the above expression. The vertex functions, $Y_2(k_l,p_m;q_n)$ and $Y_3(k_l,p_m,q_n)$, are functions of the singlet-condensation ($\bar{s}$), the Bogoliubov transformation matrix of $\mathcal{H}_{MF}$, the coupling constants ($J$'s, $D_1$, $\Gamma_1$), and the lattice vectors (${\bf R_{b,c}}$). The vertex function $Y_3(k_l,p_m,q_n)$ is totally symmetric whereas the other $Y_2(k_l,p_m;q_n)$ is symmetric only for the first two arguments: $Y_2(k_l,p_m;q_n) = Y_2(p_m,k_l;q_n)$.

\subsection{Green's function formalism}

It is convenient to recast the total Hamiltonian as follows.
\begin{equation}
 \mathcal{H}=\mathcal{H}_0+\mathcal{V}, \label{eq:H_manybody}
\end{equation}
with the kinetic part $\mathcal{H}_0=\mathcal{H}_{MF}$ (from $\mathcal{H}_{quad}+\mathcal{H}_{quartic})$ and the interaction part $\mathcal{V}=\mathcal{H}_{cubic}$.
We now define the triplon Green's function:
\begin{eqnarray}
 {\bf G}({\bf k},t)= -i \langle T \left[ \Gamma({\bf k},t) \Gamma^{\dagger}({\bf k},0) \right] \rangle ,
\end{eqnarray}
where the average $\langle \cdots \rangle$ is taken for the ground state of $\mathcal{H}$ with the time-ordering operator $T$. Here, we set $\hbar=1$, and $\Gamma({\bf k},t) = e^{i\mathcal{H}t} \Gamma({\bf k}) e^{-i\mathcal{H}t}$ with $\Gamma({\bf k})=[\gamma_1({\bf k}),\cdots,\gamma_{6}({\bf k})|\gamma_1^{\dagger}(-{\bf k}),\cdots,\gamma_{6}^{\dagger}(-{\bf k})]^T$. The Green's function is written as the following 12$\times$12 matrix:
\begin{equation}
 {\bf G}({\bf k},t)= 
 \left[
 \begin{array}{c|c}
  {\bf G}^{11}({\bf k},t) & {\bf G}^{12}({\bf k},t)
  \\
  \hline
  {\bf G}^{21}({\bf k},t) & {\bf G}^{22}({\bf k},t)
 \end{array}
 \right],
 \label{eq:G-fun}
\end{equation}
with the normal Green's function parts ${\bf G}^{11,22}$ and the anomalous function parts ${\bf G}^{12,21}$ as the 6$\times$6 submatrices.
The triplon self energy has the same matrix structure: 
\begin{equation}
 \boldsymbol{\Sigma} ({\bf k},t) = 
 \left[
 \begin{array}{c|c}
  \boldsymbol{\Sigma}^{11} ({\bf k},t) & \boldsymbol{\Sigma}^{12} ({\bf k},t)
  \\
  \hline
  \boldsymbol{\Sigma}^{21} ({\bf k},t) & \boldsymbol{\Sigma}^{22} ({\bf k},t)
 \end{array}
 \right].
\end{equation}

The Green's function and the self energy are related by the Dyson equation:
\begin{equation}
 {\bf G}({\bf k},\omega) = {\bf G}_0({\bf k},\omega) + {\bf G}_0({\bf k},\omega) \boldsymbol{\Sigma} ({\bf k},\omega) {\bf G}({\bf k},\omega).
 \label{eq:Dyson1}
\end{equation}
In the momentum-frequency space, the bare Green's function ${\bf G}_0({\bf k},\omega)$ for the Hamiltonian $\mathcal{H}_0$ is a diagonal matrix with the matrix elements
\begin{equation}
 \left[{\bf G}_0^{11}({\bf k},\omega)\right]_{mn} = \frac{\delta_{mn}}{\omega-\omega_{m}({\bf k})+i\eta},
\end{equation}
where $\omega_{m}({\bf k})~(m=1,\cdots,6)$ are the triplon eigenvalues obtained from $\mathcal{H}_{MF}$ and $\eta=0^+$. The other piece of the diagonal elements is given by the relationship ${\bf G}_0^{22}({\bf k},\omega) = {\bf G}_0^{11}(-{\bf k},-\omega)$.
Using the Dyson equation, we can express the full Green's function in terms of the bare Green's function and the self energy. For example, one can obtain
\begin{equation}
 \begin{array}{l}
 ({\bf G}^{11})^{\textup{-}1}=({\bf G}_0^{11})^{\textup{-}1}-\boldsymbol{\Sigma}^{11}-\boldsymbol{\Sigma}^{12}[({\bf G}_0^{22})^{\textup{-}1}-\boldsymbol{\Sigma}^{22}]^{\textup{-}1} \boldsymbol{\Sigma}^{21},
 \\
 \\
 {\bf G}^{21}= [({\bf G}_0^{22})^{\textup{-}1}-\boldsymbol{\Sigma}^{22}]^{\textup{-}1}\boldsymbol{\Sigma}^{21}{\bf G}^{11},
 \end{array}
 \label{eq:Dyson2}
\end{equation}
and similarly for ${\bf G}^{22,12}$.

\subsection{One-loop self energy}

\begin{figure}[t]
\includegraphics[width=0.9\linewidth]{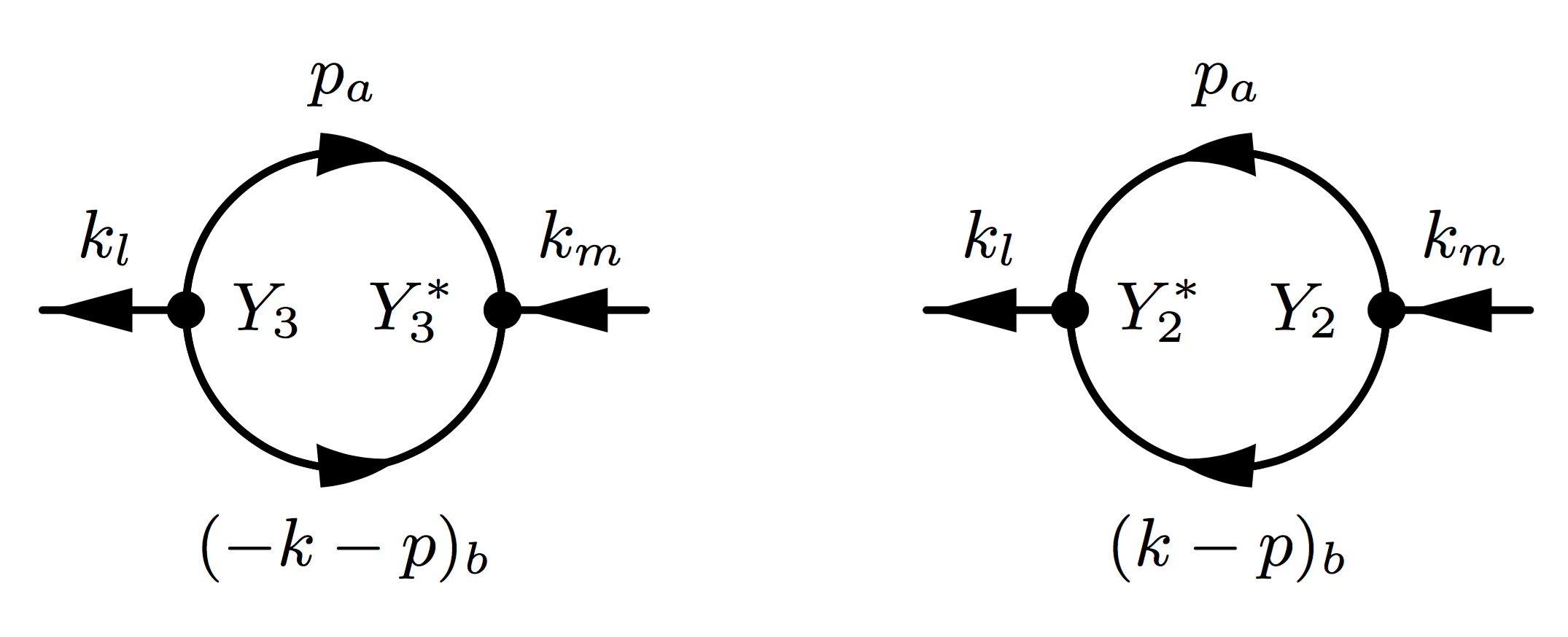}
\caption{One-loop diagrams for the self energy $\boldsymbol{\Sigma}^{11}(k)$.\label{fig:one-loop}}
\end{figure}

Let us consider one-loop self-energy correction. One-loop diagrams are drawn in Fig. \ref{fig:one-loop} for the part $\boldsymbol{\Sigma}^{11}(k)$ of the self energy. The diagrams are translated into the following equation:
\begin{eqnarray}
 &&
 [\boldsymbol{\Sigma}^{11}(k)]_{mn}
 \nonumber\\
 &=&
 \frac{1}{2}
 \sum_{\bf p} \sum_{a,b} 
 \frac{Y_2^*(p_a,(k\textup{-}p)_b;k_m) Y_2(p_a,(k\textup{-}p)_b;k_n)}
 {k_0-\omega_b(-{\bf k}+{\bf p})-\omega_a({\bf p})+i\eta}
 \nonumber\\
 &+&
 \frac{1}{2}
 \sum_{\bf p} \sum_{a,b} 
 \frac{Y_3(p_a,(\textup{-}k\textup{-}p)_b,k_m) Y_3^*(p_a,(\textup{-}k\textup{-}p)_b,k_n)}
 {-k_0-\omega_b({\bf k}+{\bf p})-\omega_a({\bf p})+i\eta}.
 ~~~~~
\end{eqnarray}
Here, we are using the abbreviated notation $k=(k_0,{\bf k})$ with $k_0$ and {\bf k} being the frequency and the momentum, respectively, and $m,n,a,b~(=1,\cdots,6)$ are triplon band indices. Besides the two diagrams in the figure, there is one more possible diagram having the vertices $Y_2$ and $Y_2^*$. However, it vanishes with no contribution to the self energy. Other parts of the self energy can be calculated in similar ways.

\subsection{Spectral function}

The Green's function can be calculated by inserting the one-loop self energy into the Dyson equation [Eq. (\ref{eq:Dyson1}) or (\ref{eq:Dyson2})]. We will extract information about the triplon modes by computing the spectral weight function of the Green's function. For positive frequencies of our interest, the spectral function is defined as
\begin{eqnarray}
 A({\bf k},\omega > 0 ) = - \frac{1}{\pi} \textup{Im} \left\{ \textup{tr}[{\bf G}({\bf k},\omega)] \right\} .
 \label{eq:spectral}
\end{eqnarray}
As a simple example, one can check that $A_0({\bf k},\omega) = \sum_{n=1}^{6} \delta[\omega-\omega_n({\bf k})]$ for the noninteracting Hamiltonian $\mathcal{H}_0$. With the triplon interaction $\mathcal{V}$, the delta-function peaks representing the bare triplon modes are modified into finite-size peaks having renormalized energy and nonzero width. If the peak is still well-defined with large height and narrow width, the associated triplon mode survives as a quasiparticle with a finite life time. In the next section, we analyze the triplon modes and their decay processes by looking at the peak structures of the spectral function.

\subsection{Results\label{sec:result}}

\begin{figure*}
\includegraphics[width=0.43\linewidth,angle=270]{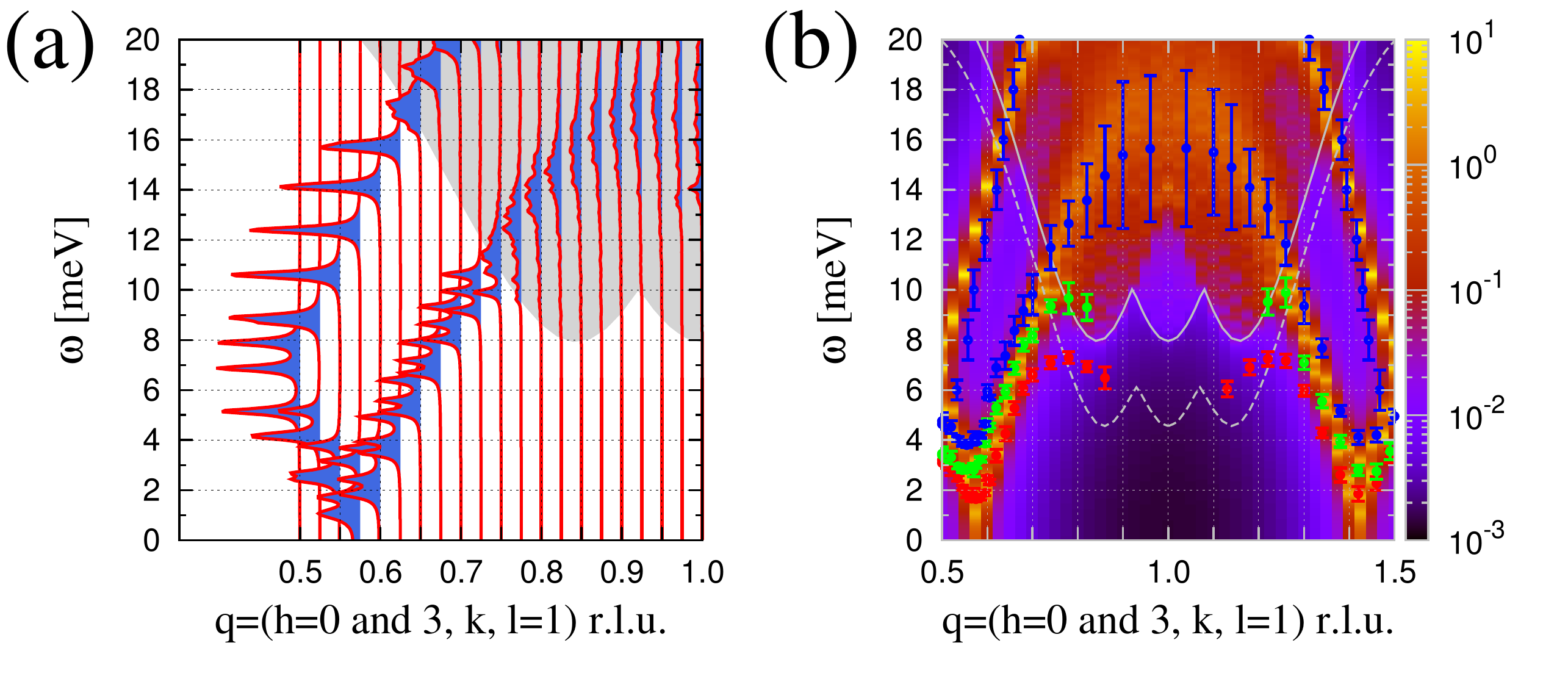}
\caption{(Color online) Spectral function, $A({\bf q},\omega)$, in the interacting triplon theory.
The spectral function is computed for the coupling constants in Eq. (\ref{eq:coupling_const2}) and plotted in the two different styles, (a) with lines and (b) with a color map.
It is compared with the neutron scattering data of Ref. [\onlinecite{2015_Plumb}] (denoted with color dots).
The gray shaded region in (a) and the solid gray line in (b) denote the multi-triplon continuum computed with Eq. (\ref{eq:two-triplon}) and the bare triplon dispersions from $\mathcal{H}_0(=\mathcal{H}_{MF})$.
The dashed gray line represents the lower boundary of the continuum obtained in the quadratic theory.
In the plot (b), the triplon continuum is partially revealed by two-triplon states generated by the triplon fusion channels in $\mathcal{V}(=\mathcal{H}_{cubic})$.
The two-triplon states are spread down to the solid gray line with weak intensities in the plot.
\label{fig:spectral}}
\end{figure*}

Here we present the results of the interacting triplon theory. We control the coupling constants such that the quasiparticle peak structures in the spectral function $A({\bf q},\omega)$ match the neutron scattering data.
The resulting renormalized coupling constants are then obtained as follows.
\begin{equation}
\begin{array}{l}
J_1=J_2=J'_2=J_4=10~\textup{meV},
\\
J_3=0.2J_1,
\\
D_1^a=D_1^b=0.3J_1,
\\
\Gamma_1^{ab}=\Gamma_1^{ba}=0.045J_1
\end{array}
\label{eq:coupling_const2}
\end{equation}
Compared to the previous estimation in Eq. (\ref{eq:coupling_const}), the Heisenberg couplings have been increased and the anisotropic couplings, $D_1$ and $\Gamma_1$, have been somewhat reduced with the inclusion of the triplon interactions. Such reductions of the anisotropic couplings reflect the fact that the $D_1$ couplings have sizeable energy renormalization effects contributing to the energy splittings around the spin gap.

The spectral function, $A({\bf q},\omega)$, calculated for the coupling constants [Eq. (\ref{eq:coupling_const2})], is plotted in Fig. \ref{fig:spectral} with the two different styles, (a) line and (b) color map. In Fig. \ref{fig:spectral} (a), we find that the triplon phenomenology mentioned before is well captured in the spectral function. First of all, the spin gap is found at $k=0.575$ with the energy, $\Delta_{th}\simeq 1.1\textup{meV}$, which is comparable to the measured value, $\Delta_{ex}= 1.67\textup{meV}$. Around the spin gap, three separated triplon modes are found with the energy splittings of $\Delta \omega_{th} \sim 2$ meV consistently with the experimental result, $\Delta \omega_{ex}=2\sim3\textup{meV}$. The energy splittings around the spin gap are mainly the energy renormalization effects of the $D_1$ couplings among the anisotropic couplings in Eq. (\ref{eq:coupling_const2}). One can notice this from the comparable sizes of $\Delta \omega_{ex}$ and $D_1^{a,b}(=3\textup{meV})$, and the small ratio $\Gamma_1/D_1=0.15$. 

Next, moving inside the triplon continuum region (gray), the triplon modes undergo substantial decay processes as shown in significantly broadened quasiparticle peak structures. Remarkably, the triplon continuum predicted in the bond operator theory matches well the region where the decay processes are found in the neutron scattering experiments. This can be seen in comparison with the experimental data in Fig. \ref{fig:spectral} (b). The second (green) and third (blue) lowest triplon modes observed in the experiments disappear or lose its prominence above the solid gray line that represents the lower boundary of the triplon continuum calculated with $\mathcal{H}_{MF}$.
For comparison, we also show the lower boundary obtained in the quadratic theory (dashed gray line) in the same plot.
The decay processes are mainly caused by the $D_1$ couplings with a minor contribution from the $J_3$ coupling (see Sec. \ref{sec:decay_channel}). It is confirmed by conducting computations with one of the two couplings being turned off.

\subsection{Discussions}

Among several experimental features in the triplon dynamics, the downward bending of the lowest triplon energy dispersion [red dots in Fig. \ref{fig:spectral} (b)] is not clearly explained in the current theory. For the bended part of the dispersion, we can think of two possibilities: (i) two-triplon bound state and (ii) the level repulsion by the triplon-continuum. The former idea naturally arises by noting that the bended part has a similar shape to that of the gray line. Although the bended part disappears at certain momentum, it could be a matrix element effect. 
Such two-triplon state contributions to the spin structure factor have been observed in the compound SrCu$_2$(BO$_3$)$_2$.\cite{2004_Knetter,2005_Aso}

On the other hand, the bended dispersion could be an effect of the level repulsion by the continuum as proposed in Ref. [\onlinecite{2015_Plumb}]. In our theory, the effects of the couplings between the single- and two-triplon excitations were implemented via the one-loop self energy in the single-triplon propagator; the level repulsion effect appears to be rather small at the one-loop level. To investigate the full level repulsion effect, it may be necessary to take into account higher order corrections or consider the single- and two-triplon sectors on equal footing. 

In both cases, these considerations would provide a unique opportunity to investigate interesting multi-particle dynamics caused by anisotropic spin interactions. 
We leave this problem as an interesting subject of future study.

\begin{table}
\begin{ruledtabular}
\begin{tabular}{cc}
Ref. [\onlinecite{2010_Tsirlin}] & This work
\\
\hline
Magnetic susceptibility & Neutron scattering
\\
(poly-crystal) & (single-crystal)
\\
\hline
Exact diagonalization & Bond operator theory
\\
(Heisenberg model) & (generic model in Eq. (\ref{eq:spin_model}))
\\
\hline
$J_1$=140 K$\simeq$12 meV & $J_1=J_2=J'_2=J_4=10$~\textup{meV}
\\
$J_2=J_1$ & $J_3=0.2J_1$
\\
$J'_2=0.5J_1$ & $D_1^a=D_1^b=0.3J_1$
\\
$J_4=0.75J_1$ & $\Gamma_1^{ab}=\Gamma_1^{ba}=0.045J_1$
\\
\end{tabular}
\end{ruledtabular}
\caption{Comparison of our work with Ref. [\onlinecite{2010_Tsirlin}]. The second and third rows indicate experimental results and theoretical approaches employed in the two works.
\label{tab:coupling-comparison}}
\end{table}

Now we comment on the coupling constants estimated from the triplon theory.
Due to many independent exchange couplings in the spin model, we considered the simplest parameter regime [Eq. (\ref{eq:coupling_const2})] that captures essential features of the neutron scattering data (see Appendix \ref{app:fitting}).
Nonetheless, we find that the overall energy scale of the parameter regime is consistent with a previous estimation in Ref. [\onlinecite{2010_Tsirlin}].
Table \ref{tab:coupling-comparison} shows our results in comparison with theirs.
Despite various differences between the two works, both results have an almost same average value of the Heisenberg couplings per a unit cell: $(4J_1+2J_2+2J'_2+2J_3+2J_4)/12 \simeq$ 8.6 meV.
A major difference is that in Ref. [\onlinecite{2010_Tsirlin}] $J'_2$ is evaluated to be much smaller than $J_2$ while in our triplon theory such a dissimilarity between $J_2$ and $J'_2$ is not crucial for describing the neutron data.
We hope this point is clarified in future studies.

The DM interactions responsible for magnetic anisotropies in BiCu$_2$PO$_6$ were estimated in our study.
Although various experimental results could be well described and understood by our theory, the magnitudes of the estimated DM interactions are quite large ($D_1^{a,b}/J_1=0.3$) compared to the values usually found in copper oxides.  
This may change once higher order contributions are taken into account in the triplon self-energy (beyond the one-loop level).
For a more accurate estimation of the coupling constants, the microscopic spin model may be directly studied with numerical techniques.
Further experimental information such as electron spin resonance (ESR) measurements will be also helpful for determining the DM interactions more precisely.\cite{1999_Nojiri,2001_Cepas,2011_Romhanyi}

\section{Summary and outlook\label{sec:conclusion}}

In this paper, we provided theoretical analysis of the rung-VBS ground state in BiCu$_2$PO$_6$ by constructing a minimal spin Hamiltonian and developing a comprehensive theory of triplon dynamics. In comparison with the neutron scattering experiment data, it is shown that the anisotropic spin interactions ($D_1$ and $\Gamma_1$) are crucial to explain the unusual quantum numbers carried by the triplons and the decay processes of the triplons to the multi-triplon continuum.

Our results would provide essential information for various ongoing studies of BiCu$_2$PO$_6$. In particular, the recent high-field experiments found a series of field-induced quantum phase transitions.\cite{2012_Kohama} Nature of the field-induced phases has not been clearly understood, in part due to the lack of the spin Hamiltonian incorporating anisotropic spin interactions. We believe that the spin Hamiltonian presented here, together with the information about the spin content of the triplons, is a good starting point for the study of the field-induced phases.

\acknowledgments

We thank Kemp Plumb and Young-June Kim for sharing their neutron scattering data on BiCu$_2$PO$_6$ and many helpful discussions.
This work was supported by the NSERC of Canada and the Canadian Institute for Advanced Research.
This research was also supported in part by Perimeter Institute for Theoretical Physics.
Research at Perimeter Institute is supported by the Government of Canada through Industry Canada
and by the Province of Ontario through the ministry of Research and Innovation.
Computations were performed on the gpc supercomputer at the SciNet HPC Consortium. SciNet is funded by: the Canada Foundation for Innovation under the auspices of Compute Canada; the Government of Ontario; Ontario Research Fund - Research Excellence; and the University of Toronto.

\appendix

\section{Anisotropic spin interactions\label{app:DM}}

The low energy spin Hamiltonian in Eq. (\ref{eq:spin_model}) can be derived from the microscopic Hubbard model consisting of the electron hopping $h$ and the on-site Coulomb repulsion $U$:
\begin{equation}
 H=\sum_{i>j} c_{i\alpha}^{\dagger} h_{ij,\alpha \beta} c_{j\beta} + U \sum_i n_{i\uparrow} n_{i\downarrow} .
\end{equation}
The electron hopping amplitude generically consists of the spin-independent ($t$) and spin-dependent (${\bf v}$) parts:
\begin{equation}
h_{ij,\alpha \beta} = t_{ij} \delta_{\alpha \beta} + i {\bf v}_{ij} \cdot \boldsymbol{\sigma}_{\alpha \beta},
\end{equation}
where $\boldsymbol{\sigma}$ are the Pauli matrices and $\alpha,\beta \in \{\uparrow,\downarrow\}$ are the spin indices. The spin-dependent hoppings have their origin in the atomic spin-orbit coupling. The corresponding hopping amplitude ${\bf v}_{ij}$ is a three-component pseudo-vector satisfying ${\bf v}_{ji}=-{\bf v}_{ij}$.
Taking the large Coulomb interaction limit ($U/h \gg 1$) with the half electron filling (one electron per site) and developing a degenerate perturbation theory, one can obtain the spin Hamiltonian in Eq. (\ref{eq:spin_model}) as a low energy effective model.\cite{DM_papers,1988_MacDonald} The coupling constants are defined in the following way.
\begin{equation}
\begin{array}{l}
J_{ij}=4t_{ij}^2/U,
\\
\\
{\bf D}_{ij}=8t_{ij}{\bf v}_{ij}/U,
\\
\\
\Gamma_{ij}^{\mu\nu}=\left(8v_{ij}^{\mu} v_{ij}^{\nu}-4\delta^{\mu\nu}{\bf v}_{ij}^2\right)/U.
\end{array}
\end{equation}
It is clear from the expressions that the Dzyaloshinskii-Moriya and anisotropic $\&$ symmetric interactions share the same origin in the spin-orbit coupling.
Their relationship in Eq. (\ref{eq:Gamma}) is obtained from the above microscopic expressions for the coupling constants.

\section{Quadratic Hamiltonian\label{app:matrix}}

The quadratic Hamiltonian takes the following form.
\begin{widetext}
\begin{eqnarray}
\mathcal{H}_{quad}
&=&
2 N_{uc} \left[ - \frac{3}{4} J_4 \bar{s}^2 + \mu (1-\bar{s}^2) \right]
+
\sum_{\bf r} \sum_{m=1}^2 \left(\frac{1}{4} J_4-\mu \right) t_{m\alpha}^{\dagger}({\bf r}) t_{m\alpha}({\bf r})
\nonumber\\
&-&
\frac{\bar{s}^2}{4} \sum_{\bf r} ( \mathcal{J}_1^{\alpha\beta} + \mathcal{J}_1^{\beta\alpha} ) [t_{1\alpha}^{\dagger}({\bf r}) t_{2\beta}({\bf r}) + t_{1\alpha}^{\dagger}({\bf r}) t_{2\beta}^{\dagger}({\bf r}) ] + \textup{H.c.}
\nonumber\\
&-&
\frac{\bar{s}^2}{4} \sum_{\bf r} ( \tilde{\mathcal{J}}_1^{\alpha\beta} + \tilde{\mathcal{J}}_1^{\beta\alpha} ) [ t_{1\alpha}^{\dagger}({\bf r}) t_{2\beta}({\bf r}-{\bf R}_{\bf b}) + t_{1\alpha}^{\dagger}({\bf r}) t_{2\beta}^{\dagger}({\bf r}-{\bf R}_{\bf b}) ] + \textup{H.c.}
\nonumber\\
&+&
\frac{\bar{s}^2}{4} \sum_{\bf r}  ( \mathcal{J}_2^{\alpha\beta} + {\mathcal{J}'}_2^{\beta\alpha} ) [ t_{1\alpha}^{\dagger}({\bf r}) t_{1\beta}({\bf r}+{\bf R}_{\bf b}) + t_{1\alpha}^{\dagger}({\bf r}) t_{1\beta}^{\dagger}({\bf r}+{\bf R}_{\bf b}) ] + \textup{H.c.}
\nonumber\\
&+&
\frac{\bar{s}^2}{4} \sum_{\bf r}  ( \mathcal{J}_2^{\alpha\beta} + {\mathcal{J}'}_2^{\beta\alpha} ) [ t_{2\alpha}^{\dagger}({\bf r}) t_{2\beta}({\bf r}-{\bf R}_{\bf b}) + t_{2\alpha}^{\dagger}({\bf r}) t_{2\beta}^{\dagger}({\bf r}-{\bf R}_{\bf b}) ] + \textup{H.c.}
\nonumber\\
&-&
\frac{\bar{s}^2}{4} \sum_{\bf r}  \mathcal{J}_3^{\alpha\beta} [ t_{1\alpha}^{\dagger}({\bf r}) t_{1\beta}({\bf r}-{\bf R}_{\bf c}) + t_{1\alpha}^{\dagger}({\bf r}) t_{1\beta}^{\dagger}({\bf r}-{\bf R}_{\bf c}) ] + \textup{H.c.}
\nonumber\\
&-&
\frac{\bar{s}^2}{4} \sum_{\bf r}  \mathcal{J}_3^{\alpha\beta} [ t_{2\alpha}^{\dagger}({\bf r}) t_{2\beta}({\bf r}+{\bf R}_{\bf c}) + t_{2\alpha}^{\dagger}({\bf r}) t_{2\beta}^{\dagger}({\bf r}+{\bf R}_{\bf c}) ] + \textup{H.c.}
\nonumber\\
&-&
\frac{\bar{s}^2}{4} \sum_{\bf r} \sum_{m=1}^2 (\mathcal{J}_4^{\alpha\beta}-J_4 \delta^{\alpha\beta}) [ t_{m\alpha}^{\dagger}({\bf r}) t_{m\beta}({\bf r}) + t_{m\alpha}^{\dagger}({\bf r}) t_{m\beta}^{\dagger}({\bf r}) ] + \textup{H.c.}
\nonumber\\
&+&
i \sum_{\bf r} \sum_{m=1}^2 g \mu_B H^{\alpha} \epsilon_{\alpha\beta\gamma} t_{m\beta}^{\dagger}({\bf r}) t_{m\gamma}({\bf r}) .
\end{eqnarray}
\end{widetext}
Here, $N_{uc}$ is the number of unit cells and ${\bf r}$ denotes a lattice vector. The coupling constant matrices denoted with $\mathcal{J}$'s are combinations of the $J$, $D$, $\Gamma$ couplings. For example,
\begin{equation}
 \mathcal{J}_1^{\alpha\beta} = J_1 \delta^{\alpha\beta} + D_1^{\gamma} \epsilon_{\gamma\alpha\beta} + \Gamma_1^{\alpha\beta} ,
 \label{eq:curlyJ_1}
\end{equation}
with $\Gamma_1$ related to $J_1$ and $D_1$ through Eq. (\ref{eq:Gamma}). The $\tilde{\mathcal{J}}_1$ matrix is defined by flipping the sign of $D_1^b$ ($D_1^b\rightarrow-D_1^b$) in the expression of $\mathcal{J}_1$. The other $\mathcal{J}$ matrices are defined in similar fashions. As mentioned in Sec. \ref{sec:quadratic_theory}, the $D_1$ couplings cancel each other in the quadratic Hamiltonian. To be specific, due to the antisymmetric nature of the DM matrix ($D_1^{\gamma} \epsilon_{\gamma\alpha\beta}$), we have $\mathcal{J}_1^{\alpha\beta} + \mathcal{J}_1^{\beta\alpha}=2 (J_1 \delta^{\alpha\beta} + \Gamma_1^{\alpha\beta})$ and similarly for $\tilde{\mathcal{J}}_1^{\alpha\beta} + \tilde{\mathcal{J}}_1^{\beta\alpha}$.

\section{Determination of exchange couplings in the quadratic theory\label{app:fitting}}

In the quadratic theory, we focus on a low energy region around the spin gap. The measured spin gap positions [$\pm{\bf q}^{\star}=\pm (0,0.425,0)$] provide a useful constraint on the Heisenberg coupling constants. The constraint can be obtained by comparing the measured value with the analytic expression for the spin gap position:
\begin{equation}
 k_{th} = \frac{1}{\pi}~\textup{cos}^{-1}\frac{J_1}{2(J_2+J_2')}.
\end{equation}
Notice that $k_{th}$ only depends on the ratio of $J_1$ to $J_2+J_2'$.\cite{2011_Lavarelo} The above expression is derived from the quadratic bond operator theory only with the Heisenberg interactions. 
Upon the inclusion of anisotropic interactions, the spin gap position remains almost the same with a slight shift, which means that the position is determined mainly by the Heisenberg interactions, $J_1$, $J_2$, $J'_2$, that generate frustration in the system.
Comparison of the theory prediction $k_{th}$ with the measured value $k_{ex}=0.425$ leads to the condition $J_1 \simeq {(J_2+J_2')}/{2}$. We choose the simplest case $J_1=J_2=J'_2$ that satisfies the condition. Additionally, we impose another condition $J_1=J_4$. These constraints are overall consistent with the previous estimation for the Heisenberg couplings in Ref. [\onlinecite{2010_Tsirlin}]. Although the latter argued that $J_2=2J'_2$ with $J_{1,2,4}$ being comparable to each other, $J_2$ and $J'_2$ are not distinguished in the bond operator Hamiltonian (as pointed out in Sec. \ref{sec:quadratic_theory}). For simplicity, we set $J_2=J'_2$. 

We find that the relatively weak coupling $J_3$ should be antiferromagnetic ($J_3>0$) to reproduce experimentally observed curvatures of the triplon dispersions at $\pm{\bf q}^{\star}$. 

As for the DM vectors, at least two components along different directions are necessary in order to completely break the spin rotation symmetry as observed in the neutron scattering experiments. $D_1^a$ and $D_1^b$ (and associated $\Gamma_1$) are found to be enough to describe low energy triplon physics as shown in Sec. \ref{sec:triplon_dispersions} and \ref{sec:triplon_magnetic_behaviors}.

\section{Mean-field decoupling\label{app:MF}}

We explain the mean-field decouplings of the quartic terms in $\mathcal{H}_{quartic}$. The quartic terms of the spin interactions at the link $ij$ are given as follows.
\begin{equation}
 \mathcal{J}_{ij}^{\alpha\beta} S_i^{\alpha} S_j^{\beta}
 \rightarrow
 -\frac{1}{4}
 \mathcal{J}_{ij}^{\alpha\beta}
 \epsilon_{\alpha\mu\nu} \epsilon_{\beta\rho\sigma}
 t_{i\mu}^{\dagger} t_{i\nu} t_{j\rho}^{\dagger} t_{j\sigma} ,
\end{equation}
where $\mathcal{J}_{ij}^{\alpha\beta}$ is a coupling constant matrix containing all the couplings ($J,~D,~\Gamma$) at the link $ij$. The quartic term is decoupled with two mean-field channels in the following way.
\begin{eqnarray}
&&
t_{i\mu}^{\dagger} t_{i\nu} t_{j\rho}^{\dagger} t_{j\sigma}
\nonumber\\
& \rightarrow &
\frac{1}{2} 
\left[
Q_{ij}^{\nu\sigma} t_{i\mu}^{\dagger} t_{j\rho}^{\dagger}
+
Q_{ij}^{*\mu\rho} t_{i\nu} t_{j\sigma}
-
Q_{ij}^{*\mu\rho} Q_{ij}^{\nu\sigma}
\right]
\nonumber\\
&+&
\frac{1}{2} 
\left[
P_{ij}^{\nu\rho} t_{i\mu}^{\dagger} t_{j\sigma}
+
P_{ij}^{*\mu\sigma} t_{i\nu} t_{j\rho}^{\dagger}
-
P_{ij}^{*\mu\sigma} P_{ij}^{\nu\rho}
\right],
\end{eqnarray}
where the mean-field parameters are defined as
\begin{equation}
\begin{array}{l}
Q_{ij}^{\nu\sigma}=\langle t_{i\nu} t_{j\sigma} \rangle ,
\\
P_{ij}^{\nu\rho}=\langle t_{i\nu} t_{j\rho}^{\dagger} \rangle .
\end{array}
\end{equation}
Although we used the site notation $i,j$ in the above for notational simplicity, in practice the computation is done with dimer indices, {\it e.g.} $i=(m,a)$ with the dimer index $m~(=1,2)$ and the spin index $a~(=L,R)$ in the dimer. At each inter-dimer interaction link, there are two mean-field parameters, $Q$ and $P$, each of which is represented by a 3$\times$3 matrix.

\section{Triplon decay and fusion channels\label{app:decay}}

Cubic terms in the bond operator theory survive when the system has couplings that are not invariant under the global site interchange at each dimer. Under the site interchange, the bond operators, based on the spin-singlet and spin-triplet basis, transform as
\begin{eqnarray}
 \left(
 \begin{array}{c}
  s
  \\
  t_{\alpha}
 \end{array}
 \right)
 \rightarrow 
 \left(
 \begin{array}{c}
  -s
  \\
  t_{\alpha}
 \end{array}
 \right)
\end{eqnarray}
with $\alpha=x,y,z$.
Since each cubic term contains one $s$-boson and three $t$-bosons, the cubic term changes its sign under the site interchange. If all the couplings in the system are invariant under the site interchange, then the cubic terms vanish. However, the spin model Eq. (\ref{eq:spin_model}) has various couplings that can generate the cubic terms. We list below such couplings with the conditions for the nonzero cubic terms.
\begin{itemize}
 \item $J_2$ and $J'_2$ when $J_2 \ne J'_2$.
 \item $J_3$.
 \item ${\bf D}_1=(D_1^a,D_1^b,D_1^c)$.
 \item ${\bf D}_2=(D_2^a,0,D_2^c)$ and ${\bf D}'_2=(D_2^{'a},0,D_2^{'c})$\\when ${\bf D}_2 \ne -{\bf D}'_2$.
 \item ${\bf D}_3=(0,D_3^b,0)$.
 \item ${\bf D}_4=(0,D_4^b,0)$.
\end{itemize}
For simplicity, we have presented only the $J$ and $D$ couplings without the less important $\Gamma$ couplings in the decay processes.


\begin{thebibliography}{99}

\bibitem{2010_Balents} L. Balents, Nature (London) {\bf 464}, 199 (2010).

\bibitem{2007_Koteswararao} B. Koteswararao, S. Salunke, A. V. Mahajan, I. Dasgupta, and J. Bobroff, Phys. Rev. B {\bf 76}, 052402 (2007).


\bibitem{2009_Mentre} O. Mentr\'e, E. Janod, P. Rabu, M. Hennion, F. Leclercq-Hugeux, J. Kang, C. Lee, M.-H. Whangbo, and S. Petit, Phys. Rev. B {\bf 80}, 180413 (2009).


\bibitem{2010_Alexander} L. K. Alexander, J. Bobroff, A. V. Mahajan, B. Koteswararao, N. Laflorencie, and F. Alet, Phys. Rev. B {\bf 81}, 054438 (2010).

\bibitem{2010_Tsirlin} A. A. Tsirlin, I. Rousochatzakis, D. Kasinathan, O. Janson, R. Nath, F. Weickert, C. Geibel, A. M. L\"auchli, and H. Rosner, Phys. Rev. B {\bf 82}, 144426 (2010).

\bibitem{2011_Lavarelo} A. Lavar{\'e}lo, G. Roux, and N. Laflorencie, Phys. Rev. B {\bf 84}, 144407 (2011).

\bibitem{2012_Kohama} Y. Kohama, S. Wang, A. Uchida, K. Prsa, S. Zvyagin, Y. Skourski, R. D. McDonald, L. Balicas, H. M. R{\o}nnow, C. R\"uegg, and M. Jaime, Phys. Rev. Lett. {\bf 109}, 167204 (2012).

\bibitem{2013_Choi} K.-Y. Choi, J. W. Hwang, P. Lemmens, D. Wulferding, G. J. Shu, and F. C. Chou, Phys. Rev. Lett. {\bf 110}, 117204 (2013).

\bibitem{2013_Casola} F. Casola, T. Shiroka, A. Feiguin, S. Wang, M. S. Grbi\'c, M. Horvati\'c, S. Kr\"amer, S. Mukhopadhyay, K. Conder, C. Berthier, H.-R. Ott, H. M. R{\o}nnow, C. R\"uegg, and J. Mesot, Phys. Rev. Lett. {\bf 110}, 187201 (2013).

\bibitem{2013_Sugimoto} T. Sugimoto, M. Mori, T. Tohyama, and S. Maekawa, Phys. Rev. B {\bf 87}, 155143 (2013).

\bibitem{2013_Plumb} K. W. Plumb, Z. Yamani, M. Matsuda, G. J. Shu, B. Koteswararao, F. C. Chou, and Y.-J. Kim, Phys. Rev. B {\bf 88}, 024402 (2013).

\bibitem{2015_Plumb} K. W. Plumb, K. Hwang, Y. Qiu, L. W. Harriger, G. E. Granroth, G. J. Shu, F. C. Chou, C. R\"uegg, Y. B. Kim, Y.-J. Kim, Nat. Phys. {\bf 12}, 224 (2016).

\bibitem{1990_Sachdev} S. Sachdev and R. N. Bhatt, Phys. Rev. B {\bf 41}, 9323 (1990).

\bibitem{1994_Gopalan} S. Gopalan, T. M. Rice, and M. Sigrist, Phys. Rev. B {\bf 49}, 8901 (1994).

\bibitem{2004_Matsumoto} M. Matsumoto, B. Normand, T. M. Rice, and M. Sigrist
Phys. Rev. B {\bf 69}, 054423 (2004).

\bibitem{DM_papers} I. Dzyaloshinskii, J. Phys. Chem. Solids {\bf 4}, 241 (1958); T. Moriya, Phys. Rev. {\bf 120}, 91 (1960).

\bibitem{Goodenough_Kanamori} J. B. Goodenough, Phys. Rev. {\bf 100}, 564 (1955); J. Kanamori, J. Phys. Chem. Solids {\bf 10}, 87 (1959).

\bibitem{extrapolation} As we approach the critical points, it takes a longer and longer time to obtain a converged solution of the saddle point equations [Eq. (\ref{eq:saddle-point-eq})]. Rather than solving the equations directly around the critical points, we estimate the critical fields by extrapolation.

\bibitem{2013_Zhitomirsky} M. E. Zhitomirsky and A. L. Chernyshev, Rev. Mod. Phys. {\bf 85}, 219 (2013), and references therein.

\bibitem{2006_Zhitomirsky} M. E. Zhitomirsky, Phys. Rev. B {\bf 73}, 100404(R) (2006).

\bibitem{2003_Fetter} A. L. Fetter and J. D. Walecka, {\it Quantum Theory of Many-Particle Systems} (Dover Publications, 2003).


\bibitem{2004_Knetter} C. Knetter and G. S. Uhrig, Phys. Rev. Lett. {\bf 92}, 027204 (2004).

\bibitem{2005_Aso} N. Aso, H. Kageyama, K. Nukui, M. Nishi, H. Kadowaki, Y. Ueda, and K. Kakurai, J. Phys. Soc. Jpn. {\bf 74}, 2189 (2005).

\bibitem{1999_Nojiri} H. Nojiri, H. Kageyama, K. Onizuka, Y. Ueda, and M. Motokawa, J. Phys. Soc. Jpn. {\bf 68}, 2906 (1999).

\bibitem{2001_Cepas} O. C{\'e}pas, K. Kakurai, L. P. Regnault, T. Ziman, J. P. Boucher, N. Aso, M. Nishi, H. Kageyama, and Y. Ueda, Phys. Rev. Lett. {\bf 87}, 167205 (2001).

\bibitem{2011_Romhanyi} J. Romh{\'a}nyi, K. Totsuka, and K. Penc, Phys. Rev. B {\bf 83}, 024413 (2011).

\bibitem{1988_MacDonald} A. H. MacDonald, S. M. Girvin, and D. Yoshioka, Phys. Rev. B {\bf 37}, 9753 (1988).




\end{thebibliography}
\end{document}